\documentclass[twocolumn, twocolappendix]{aastex7}

\usepackage{amsmath, mathtools, amsthm, amssymb}
\usepackage{enumitem}
\usepackage{booktabs}
\usepackage{makecell}
\usepackage[toc,page]{appendix}
\hypersetup{colorlinks=true, allcolors=blue}
\usepackage{soul}
\usepackage{graphicx,color}
\usepackage{hyperref}
\usepackage{comment}
\usepackage{chngcntr}
\usepackage{appendix}
\graphicspath{ {./} }
\usepackage{xcolor}

\newcommand{\msol}{\ensuremath{M_{\odot}}}
\newcommand{\NC}{\mbox{\sc \small NewCluster}}
\newcommand{\chemcomp}{${\rm MgFeSiO}_{\rm 4}$}
\newcommand{\mytilde}{\raise.19ex\hbox{$\scriptstyle\sim$}}
\newcommand{\gm}{$G\text{\textendash} M_{20}$}

\usepackage[super]{nth}

\usepackage[normalem]{ulem}
\usepackage{threeparttable}

\submitjournal{ApJ}
\shortauthors{Byun et al.}
\shorttitle{Dust Models Shape High-$z$ Galaxy Morphology}

\begin{document}

\title{How Dust Models Shape High-$z$ Galaxy Morphology: Insights from the NewCluster Simulation}

\author[0009-0007-8611-3813]{Gyeong-Hwan Byun}\email{byunastro@yonsei.ac.kr}
\affil{Department of Astronomy and Yonsei University Observatory, Yonsei University, Seoul 03722, Republic of Korea}

\author[0000-0002-0858-5264]{J. K. Jang}\email{starbrown816@yonsei.ac.kr}
\affil{Department of Astronomy and Yonsei University Observatory, Yonsei University, Seoul 03722, Republic of Korea}

\author[0009-0009-4086-7665]{Zachary P. Scofield}\email{zpscofield15@gmail.com}
\affil{Department of Astronomy and Yonsei University Observatory, Yonsei University, Seoul 03722, Republic of Korea}

\author[0009-0009-4676-7868]{Eunmo Ahn}\email{eunmo.ahn.ast@gmail.com}
\affil{Department of Astronomy and Yonsei University Observatory, Yonsei University, Seoul 03722, Republic of Korea}

\author[0000-0002-3930-2757]{Maarten Baes}\email{maarten.baes@ugent.be}
\affil{Sterrenkundig Observatorium, Universiteit Gent, Krijgslaan 281, B-9000 Gent, Belgium}

\author[0000-0003-0225-6387]{Yohan Dubois}\email{dubois@iap.fr}
\affil{Institut d’ Astrophysique de Paris, Sorbonne Universités, et CNRS, UMP 7095, 98 bis bd Arago, 75014 Paris, France}

\author[0000-0001-9939-713X]{San Han}\email{san.han@iap.fr}
\affil{Institut d’ Astrophysique de Paris, Sorbonne Universités, et CNRS, UMP 7095, 98 bis bd Arago, 75014 Paris, France}

\author[0000-0002-1270-4465]{Seyoung Jeon}\email{syj3514@yonsei.ac.kr}
\affil{Department of Astronomy and Yonsei University Observatory, Yonsei University, Seoul 03722, Republic of Korea}

\author[0000-0002-4391-2275]{Juhan Kim}\email{kjhan@kias.re.kr}
\affil{Korea Institute of Advanced Studies, 85 Hoegiro, Dongdaemun-gu, Seoul 02455, Republic of Korea}

\author[0000-0003-0695-6735]{Christophe Pichon}\email{pichon@iap.fr}
\affil{Institut d’ Astrophysique de Paris, Sorbonne Universités, et CNRS, UMP 7095, 98 bis bd Arago, 75014 Paris, France}
\affil{Kyung Hee University, Dept. of Astronomy \& Space Science, Yongin-shi, Gyeonggi-do 17104, Republic of Korea}

\author[0000-0002-0184-9589]{Jinsu Rhee}\email{jinsu.rhee@gmail.com}
\affil{Department of Astronomy and Yonsei University Observatory, Yonsei University, Seoul 03722, Republic of Korea}
\affil{Institut d’ Astrophysique de Paris, Sorbonne Universités, et CNRS, UMP 7095, 98 bis bd Arago, 75014 Paris, France}
\affil{Korea Astronomy and Space Science Institute, 776, Daedeokdae-ro, Yuseong-gu, Daejeon 34055, Republic of Korea}

\author[0000-0001-6535-1766]{Francisco Rodr\'iguez Montero}\email{currodri@gmail.com}
\affil{Institut d’ Astrophysique de Paris, Sorbonne Universités, et CNRS, UMP 7095, 98 bis bd Arago, 75014 Paris, France}

\author[0000-0002-4556-2619]{Sukyoung K. Yi}\email{yi@yonsei.ac.kr}
\affil{Department of Astronomy and Yonsei University Observatory, Yonsei University, Seoul 03722, Republic of Korea}

\begin{abstract}
Dust plays a pivotal role in shaping the observed morphology of galaxies. 
While traditional cosmological simulations often assume a fixed dust-to-gas (DTG) or dust-to-metal (DTM) mass ratio to model dust effects, recent advancements have enabled on-the-fly (OTF) dust modeling that captures the spatial and temporal evolution of dust.
In this work, we investigate the impact of dust modeling on galaxy morphology using the \NC\ simulation, which implements a detailed OTF dust model. 
We generate mock images of \NC\ galaxies under both OTF and fixed DTM models using the radiative transfer code \mbox{\sc \small SKIRT}, and compare their morphology to JWST observations. 
We measure morphology indices and use the \gm\ test to classify galaxies.
We find that the OTF galaxy models exhibit brighter centers and more pronounced bulges than those of the fixed DTM models, resulting in a lower late-type galaxy (LTG) fraction, particularly at high redshifts. This central brightening is linked to a phenomenon we refer to as the \textit{DTM cavity}, a localized depression in the DTM ratio driven by intense bulge starbursts. Our results highlight the importance of modeling dust evolution in a physically motivated manner, as fixed DTM models fail to capture key morphological features.
\end{abstract}

\section[]{Introduction}
\label{sec:intro}
Morphology is one of the most fundamental properties of galaxies, providing a first impression of their nature. Since \citet{Hubble_1926} first proposed a morphological classification scheme, systematic studies of galaxy morphology have been conducted from various perspectives \citep[e.g.,][]{1959HDP....53..275D, vandenBergh_1960, Sandage_1961, Dressler_1994, Conselice_2003, lotz2008gm20}. These studies have revealed that morphology not only characterizes intrinsic structural features but also reflects the complex interplay between galaxies and their surrounding environments, as well as the impact of internal feedback mechanisms.

In the framework of hierarchical structure formation \citep[e.g.,][]{WhiteRees_1978, Davis_1985, Kauffmann_1993}, galaxies are understood to form and evolve within dark matter halos. 
Late-type galaxies (LTG), such as disk galaxies, are expected to preferentially form in halos with higher angular momentum \citep{Fall_efstathiou_1980, MoMaoWhite_1998, Bosch_2002, Agertz_2011, Somerville_2015}, and gradually stabilize through disk settling mechanisms \citep{Kassin_2012,  McCluskey_2024}. However, over cosmic time, these disk galaxies may transform into spheroidal, dispersion-dominated galaxies. While mergers have often been invoked to explain these transformations, recent studies have emphasized that the evolutionary pathways are considerably more complex, involving multiple interacting physical processes \citep{Negroponte_1983, Moore_1996, Sales_2012, Martin_2018}.

Therefore, accurately understanding the evolution of galaxy morphology is crucial for testing galaxy formation theories. Since morphological transformations occur early in the history of structure formation \citep{Leste_2024, lee_2024}, high-redshift galaxy morphology provides key insights into the initial stages of galaxy evolution. Despite its significance, studying high-redshift galaxy morphology remains challenging due to observational limitations.
The advent of JWST has ushered in a new era, enabling high-redshift galaxy morphology to be determined in unprecedented detail \citep[e.g.,][]{Ferreira_2022, Kuhn_2024} and allowing for direct comparisons with theoretical predictions from cosmological simulations. However, accurately interpreting these observations requires careful consideration of dust, which can significantly alter the appearance of galaxies by absorbing and scattering light.

Given the difficulty of constraining dust properties observationally at high redshifts, cosmological hydrodynamic simulations provide a valuable alternative for investigating its impact on galaxy morphology.
Consequently, many studies have incorporated dust effects into simulations when producing mock observations that can be directly compared with JWST and other observational data. These efforts are crucial for bridging the gap between theory and observation.

However, traditional simulations without explicit dust evolution models have approximated dust distribution using a fixed dust-to-gas (DTG) or dust-to-metal (DTM) ratio \citep[e.g.,][]{Narayanan_2010, Somerville_2012, Hayward_2014, Lacey_2016, Camps_2016, Rodriguez_2019}. This approach assumes a uniform relationship between dust and metals, thereby simplifying the modeling process, yet it fails to capture the self-consistent nature of dust evolution \citep{inoue_2003, hirashita_2011, Aoyama_2017}.

Early studies suggested that adopting a fixed DTM ratio across all galaxies provided a reasonable approximation \citep{Dwek_1998, Zafar_2013}. However, recent research has demonstrated that DTG and DTM values exhibit significant variations depending on galaxy properties \citep{RemyRuyer_2014, Popping_2017, Devis_2019} and even within individual galaxies \citep{Mattsson_2012}. These findings highlight the need for a more realistic dust modeling approach.
Moreover, this issue becomes particularly problematic at high redshifts. Observations from JWST and ALMA suggest that the dust mass, DTG ratios, and DTM ratios in the high-redshift universe differ from those in the local universe \citep{Algera_2025}.
Consequently, the use of a fixed DTM ratio in galaxy morphology studies may introduce significant biases, limiting the ability of simulations to produce realistic mock observations. 

To overcome these limitations, a more advanced approach is needed that accurately captures the evolving nature of dust in galaxies. In response to this need, ``on-the-fly'' (OTF) dust models have been developed \citep[e.g.,][]{Bekki_2013, Hirashita15, McKinnon_2017, li2019, aoyama2020, Dubois24, choban2024a}. Unlike fixed DTM models, OTF models self-consistently track dust formation and destruction, providing a more physically motivated representation of dust properties. This advancement facilitates a more accurate assessment of how dust influences the observed morphology.

In this work, we use the \NC\ simulation, a high-resolution cosmological zoom-in simulation that incorporates detailed dust physics. The \NC\ region traces a $4\,\sigma$ overdensity, providing a representative environment for studying massive galaxies in the early universe and enabling meaningful comparisons with deep observations. With its high spatial and mass resolution (see Section~\ref{sec:sim}), \NC\ resolves internal galactic structures and allows for a direct assessment of how dust affects observed morphology.
To obtain realistic galaxy morphologies, we perform mock observations using \mbox{\sc \small SKIRT} \citep{SKIRT2015, camps2020SKIRT9}. We investigate how the choice of dust modeling—specifically, OTF versus fixed DTM ratio models—affects simulated galaxy morphology at high redshift. We then discuss how these morphological differences arise from a physical mechanism. Finally, we argue that a realistic treatment of dust is essential as cosmological simulations increasingly aim to model dust evolution in a physically motivated way.

\section[]{Simulation}
\label{sec:sim}
\subsection{NewCluster}
We utilize the \NC\ simulation \citep{nc_arXiv_2025}, a high-resolution zoom-in cosmological simulation using RAMSES-yOMP \citep{ramsesyomp}, an OpenMP version of RAMSES \citep{teyssier_2002}. The simulation begins with a periodic box of side length 100 Mpc/h and focuses on a single galaxy cluster at the center of the zoom-in region. The target cluster has a virial mass of $\sim 4.7 \times 10^{14} \msol$, with a zoom-in region radius of $3.5 R_{\rm{vir}} \sim 17.7\, \rm{Mpc/h}$ at $z = 0$. The mass resolution of dark matter (DM) in the zoom-in region is $m_{\rm{DM}} \sim 1.3\times 10^6 M_{\odot}$, and for stellar particles, $m_{\star} \sim 2\times 10^4 M_{\odot}$. The best spatial resolution varies with the scale factor of the simulation, ranging from 53 to 107 pc in physical size. \NC\ adopts the WMAP-7 cosmological parameters $(H_0=70.3\, {\rm km\, s}^{-1}\,{\rm Mpc}^{-1}, \Omega_m=0.272, \Omega_b = 0.0455, \Omega_{\rm \Lambda}=0.728, \sigma_8=0.810\ {\rm and}\ n_s=0.967)$ \citep{wmap7}.

\NC\ adopts radiative cooling and heating based on the assumption of collisional ionization equilibrium. Radiative cooling rates are calculated using the tabulations from \citet{sutherland_dopita1993}, which allow primordial gas to cool down to $10^4\, \rm{K}$. Below this temperature, metal cooling further enables gas to cool to as low as $0.1\, \rm{K}$. Radiative heating from a uniform UV background is activated at $z=10$, following the prescription of \citet{haardt_madau1996}. Additionally, as the gas becomes optically thick, it self-shields the UV background by a factor of $\exp{({-n_{\rm H}/n_{\rm shield}})}$, where $n_{\rm shield}$ is self shielding threshold density set to $0.01\, \rm H\, cm^{-3}$.

\NC\ incorporates both stellar and supermassive black hole (SMBH) physics.
Star formation occurs in gas cells with a hydrogen number density exceeding the threshold of $n_{\rm th} = 5\,\rm{cm^{-3}}$, with the star formation efficiency from \citet{federrath_2012}.
To realistically model the chemical evolution of galaxies, \NC\ includes Type II and Type Ia supernovae (SNe), as well as stellar winds, and tracks the evolution of ten chemical elements: H, D, He, C, N, O, Mg, Fe, Si, and S. 
To compute chemical enrichment, \NC\ adopts the Chabrier initial mass function (IMF) \citep{chabrier2003IMF} for each stellar particle.
Considering each stellar particle as a simple stellar population, the chemical enrichment of ten chemical elements is computed using the Starburst99 code \citep{starburst99}.
The yields and energy outputs from stellar winds are based on \citet{Schaller92, Maeder00}, and they are directly injected into the simulation grid cell containing the corresponding stellar particle.

Every stellar particle triggers SN II explosions once it reaches an age of $5\, \rm Myr$.
For SN II events, the chemical yields depend on the progenitor mass and follow the model of \citet{kobayashi2006} for progenitors in the range of $8 - 50\, M_{\odot}$. 
The yields of SNe Ia are derived from \citet{iwamoto1999}, with their frequency determined by a delay time distribution modeled as a power-law function. The feedback energy of a single SN Ia is set to $10^{51}\, \rm{erg}$. 
Both the SNe II and SNe Ia feedback are implemented using the mechanical feedback scheme of \citet{kimm2014_SNIa}.

\NC\ incorporates dual-mode active galactic nucleus (AGN) feedback, following the model of \citet{dubois2012_AGN}, where the mode is determined by the Eddington ratio. When the Eddington ratio is below 0.01, black holes release mass and energy as jets (jet mode), while above 0.01, they emit thermal energy isotropically (quasar mode). The Eddington ratio is capped at 1 to ensure it does not exceed the Eddington limit.

\NC\ has passed $z=0.8$, and a more detailed description of the simulation and some key features of galaxies in \NC\, can be found in \citep{nc_arXiv_2025}.

\subsection{Dust models}
\label{sec:nc_dust}

In \NC, the dust component is treated as a passive scalar parameter that evolves on the fly.
The dust model implementation is similar to that of \citet{Dubois24}. It should be noted that \citet{Dubois24} used idealized MW-like galaxy simulations, while \NC\ is a cosmological simulation. Any deviations caused by this difference with \citet{Dubois24} are explicitly highlighted in this section.

The dust size distribution of \NC\ follows the two-size approximation from \citet{Hirashita15}, dividing the grains into small and large size populations. 
The small and large grain populations have sizes of $5\, \rm{nm}$ and $0.1\, \mu\mathrm{m}$, respectively.
Each size population is affected differently by various dust evolution processes. Dust models with a two-size approximation are consistent with results obtained by using a full treatment of the size distribution \citep{aoyama2020}.
Since \NC\ calculates the evolution of 9 elements, it is utilized to make up the two chemical types of dust: silicate and carbonaceous. The silicate grains are assumed to consist of iron-rich olivine (\chemcomp) \citep{kemper2004, min2007}, while the carbonaceous grains are composed solely of carbon. The material densities are set to $s_{\rm C} = 2.2\, {\rm g}\,{\rm cm}^{-3}$ for the carbonaceous grains and $s_{\rm Si} = 3.3\, {\rm g}\, {\rm cm}^{-3}$ for the silicate grains.

\subsubsection{Dust formation}
\label{sec:dust_form}
The elements contributing to the chemical composition of dust are released into the interstellar medium by stellar ejecta. Some of these ejected elements condense into dust, with the fraction determined by the condensation efficiency $\delta_{i}^{k}$. Here, the superscript $k$ represents the different sources of stellar yield ($\rm k=\rm SNII,\ SNIa,\ or\ AGB$) and the subscript $i$ represents the chemical type (carbonaceous or silicate).
The values of condensation efficiencies are the same as in \citet{Dubois24}. We also adopt the condensation efficiencies from \citet{Dwek_1998} with the exception of $\delta^{\rm SNIa}_{\rm Si}$. As noted by \citet{Dwek_1998}, however, the choice of condensation efficiencies is somewhat arbitrary. Moreover, these efficiencies remain highly uncertain and may vary by up to an order of magnitude for the same objects \citep{schneider_2024}.

For low-metallicity galaxies at high redshift, dust is primarily produced by stellar ejecta, especially from SNe II. As galaxies evolve and metal enrichment progresses, dust accretion becomes the dominant formation mechanism \citep{zhukovska2014, hou2019, li2019, choban2024a}. The accretion follows: 
\begin{equation}
    \dot{M}_{{\rm{D}},i,j} = \bigg( 1-\frac{M_{{\rm{D}},i,j}}{M_{{\rm{Z}},i}} \bigg) \frac{M_{{\rm{D}},i,j}}{t_{{\rm{acc}},i,j}},
\end{equation}
where $M_{{\rm{Z}},i}$ is the total mass of metals in both gas and dust phases, while $M_{{\rm{D}},i,j}$ represents the dust mass of grain type $i$ and size $j$.
The accretion timescale follows:
\begin{equation}
\begin{aligned}
    t_{{\rm{acc}},i,j} = C_X\,\alpha^{-1} a_{{\rm 0.005},j} s_{3,i} \frac{A_X^{1/2}f_X}{Z_X} \\
    \times \left(\frac{n}{1\ {\rm H}\, {\rm cm}^{-3}}\right)^{-1} \left(\frac{T}{50\ {\rm K}}\right)^{-1/2} \rm Myr,
\end{aligned}
\end{equation}

where $\alpha$ is the sticking coefficient, $a_{0.005,j}=a_j/5\, \rm{nm}$ is the normalized grain size, $s_{3,i}=s_i/3\,\rm{g\,cm^{-3}}$ is normalized grain density, $A_X$ is the atomic number of the limiting element $X$, $f_X$ is its mass fraction in the grain chemical composition, and $Z_X$ is its mass fraction in the gas phase. 
$T$ and $n$ denote the gas temperature and density, respectively.
The constant $C_X$ takes values of 0.28, 0.34, 0.23, 0.32, or 0.42 when the limiting element is C, Mg, Fe, Si, or O, respectively.
To account for unresolved scales, we employ a subgrid accretion model that assumes the gas follows a log-normal probability distribution function in density, temperature, and metallicity for gas cells with $n > 0.1\,{\rm H\, cm^{-3}, T<10^4\, K}$ (see \citet{Dubois24} for the detailed implementation).
We do not consider the Coulomb enhancement factor, which results from electrostatic effects, in the calculation of the accretion timescale.

\subsubsection{Dust destruction}
Once the dust has been formed, it can be destroyed by SN shocks, thermal sputtering, and astration.
The amount of dust destruction, $\Delta M_{{\rm SN},j}$, is determined by $M_{100} / M_{g\rm }$, the mass fraction of gas shocked at above 100 km/s:
\begin{equation}
\Delta M_{{\rm SN},j} =\epsilon_{{\rm SN}}(a_{j})min\left(\frac{M_{\rm 100}}{M_{\rm g}},1\right)M_{{\rm D},j}.
\label{sn_destroy}
\end{equation}
The grain size-dependent destruction efficiency $\epsilon_{{\rm SN},i}(a_j)$ follows \citet{aoyama2020}:
\begin{equation}
    \epsilon_{\rm SN}(a_j) = 1-\exp\left(-\frac{\delta_{{\rm SN}}}{a_{0.1,j}}\right).
\end{equation}
Here, the destruction efficiency does not depend on the composition of the grain or the type of SNe. 
Therefore, the destruction efficiency is uniformly given by $\delta_{\rm SN} = 0.10$. However, \citet{Dubois24} use different destruction efficiency for the composition of the grain $(\delta_{\rm SN,C} = 0.10, \delta_{\rm SN,Si}=0.15)$.

In the thermal sputtering process, dust grains gain energy exceeding their binding energy through interactions with hot gas, resulting in the transfer of mass from dust to gas-phase metals. Since \NC\ does not include radiative transfer calculations, non-thermal sputtering processes are not implemented. 
The destruction rate of dust by thermal sputtering, $\dot{M}_{{\rm spu},j}$, is determined by $t_{\rm spu}$, sputtering timescale:
\begin{equation}
    \dot{M}_{{\rm spu}, j} = - M_{{\rm D},j} / t_{\rm spu}, 
    \label{sputtering}
\end{equation}
where the sputtering timescale follows \citet{tsai_1995}:
\footnote{Since it is more likely to be $\frac{M_{\rm spu}}{\dot{M_{\rm D}}}= \frac{a}{3\dot{a}}$, our thermal sputtering time scale is likely overestimated by a factor of 3 in Equation \ref{sputtering}.}
\begin{align}
    &t_{\rm spu} = \frac{a_{0.1}}{\dot{a}_{0.1}} \notag \\
    &= 0.165\, a_{0.1} \left(\frac{n}{1\, {\rm H}\, {\rm cm}^{-3}}\right)^{-1} 
    \left[\left( \frac{2 \times 10^6\,\rm K}{T}\right)^{2.5}+1 \right]\, \rm Myr.
\end{align}
Unlike \citet{Dubois24}, we do not consider thermal sputtering yields to depend on chemical grain type.

When a stellar particle forms, we enforce that the DTG ratio in the parent gas cell remains unchanged. Since gas mass is removed during star formation, the corresponding amount of dust mass is also removed and incorporated into the stellar particle. This process effectively represents astration, the destruction of dust due to star formation.

\subsubsection{Dust size evolution}
Depending on the velocity dispersion of dust in cold gas, dust grains can coagulate or shatter through direct collisions. These processes transfer mass between grains of different sizes while conserving the total dust mass.
Coagulation primarily occurs where the gas density is high ($n \sim 10^2 - 10^3\, {\rm H}\,{\rm cm}^{-3}$) and the velocity dispersion is low ($\sigma_{\rm{D,S}} \sim 0.1 - 1\ {\rm km\, s^{-1}}$) \citep{yan2004}. 
Such high densities present a challenge as they are unresolvable at the resolution of galaxy formation simulations. Therefore, \NC\ adopts a fraction of dense gas, denoted as $F$ (or ``fudge factor"), as proposed by \citet{Aoyama_2017}. For simplicity, we use $F = 0.5$ in this study. 
In this scheme, half of the gas mass has a density $n > 10^3\, \rm{H\, cm^{-3}}$ in unresolved gas cells. 
\cite{Dubois24} conducted tests with two simulations, one with $F = 1$ and the other where $n$ follows a log-normal PDF shaped by turbulence. They showed 0.2\% and 3\% fewer small grains, respectively, compared to the $F=0.5$ case. 
Therefore, the impact of the choice of $F$ is small. 

The timescale for coagulation, $t_{\rm coa}$, is calculated following \citet{Aoyama_2017}:
\begin{equation}
    t_{\rm coa} = 0.27 \frac{a_{0.005}s_{3}}{F}\left(\frac{\mathcal{D}_{\rm S}}{0.01}\right)^{-1}\left(\frac{\sigma_{\rm D,S}}{0.1\ {\rm km\, s^{-1}}}\right)^{-1}\ \rm Myr,
\end{equation}
where $\mathcal{D}_{\rm S}$ is dust-to-gas ratio of small grains. 

The shattering process used in \NC\ is identical to that of \citet{Dubois24}. They adopted the model revised by \citet{Granato2021}:
\begin{equation}
    t_{\rm sha} = 54 a_{0.1} s_{3} \left(\frac{\mathcal{D}_{\rm L}}{0.01}\right)^{-1} \left(\frac{n}{1\ {\rm H}\, {\rm cm}^{-3}}\right)^{-p_{\rm sh}}\ \rm Myr,
\end{equation}
where $p_{\rm sh} = 1$ in the gas with $n < 1\ {\rm H}\, {\rm cm}^{-3}$ and $p_{\rm sh} = 1/3$ in the gas with $1 < n < 10^3\ {\rm H}\, {\rm cm}^{-3}$.

\begin{figure}[htbp]
    \centering
	\includegraphics[width=\columnwidth]{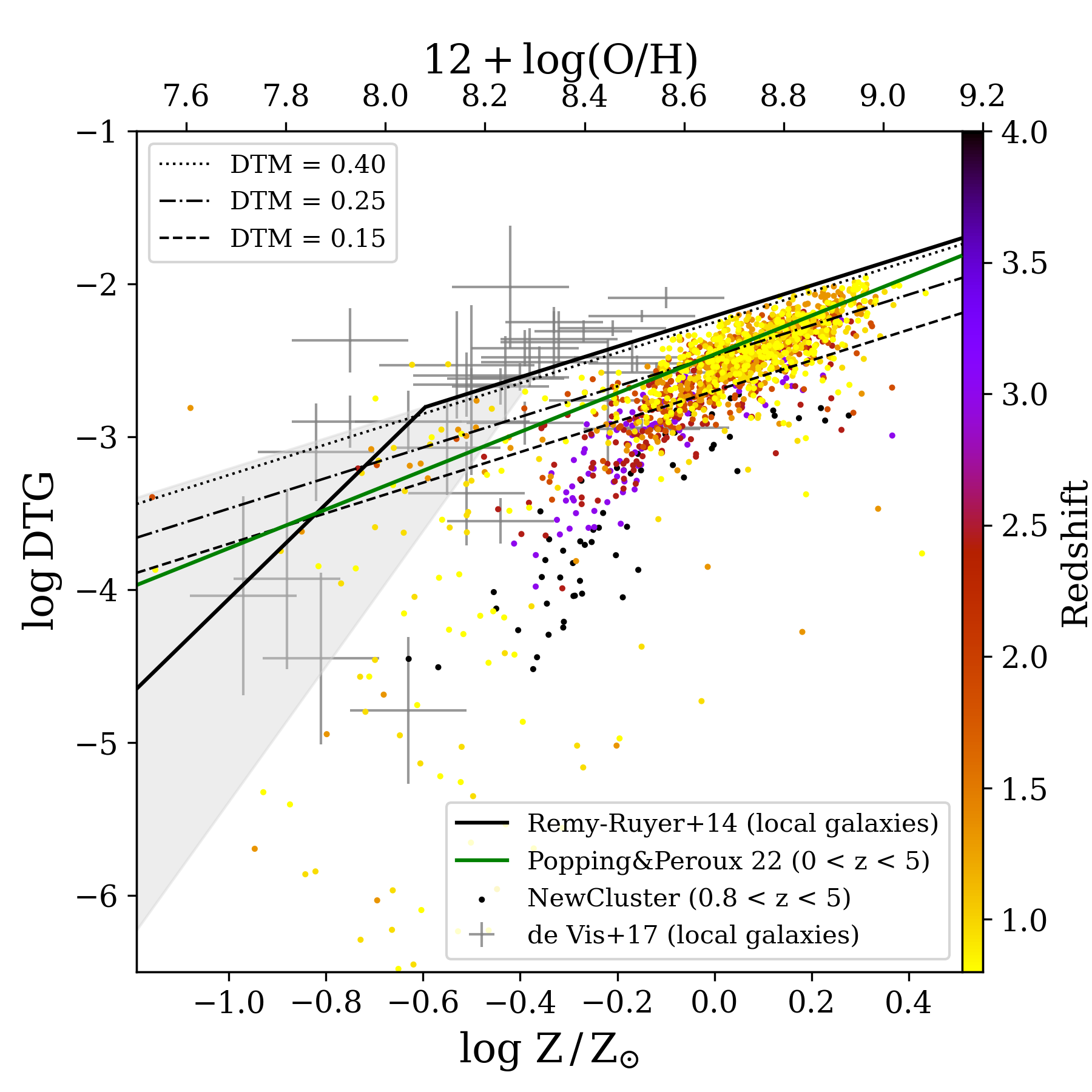}
   \caption{The relation between the DTG ratio and gas-phase metallicity for \NC\ galaxies. The color of each point indicates the redshift. The solid black line represents the broken power-law fit to observational data from \citet{RemyRuyer_2014}, and the green line represents the linear regression fit from \citet{popping_2022}.
   The shaded region denotes the uncertainty of the \citet{RemyRuyer_2014} fit. The dashed, dot-dashed, and dotted black lines correspond to constant DTM ratios of 0.15, 0.25, and 0.40, respectively. The gray crosses are the observational measurements from Table A1 and A2 in \citet{deVis_2017}.}
    \label{fig:scaling_relation}
\end{figure}

To assess the validity of the adopted dust model, we compare the simulated DTG-metallicity relation with observations. Figure~\ref{fig:scaling_relation} shows the DTG ratio as a function of gas-phase metallicity, measured within $r_{80}$, the radius enclosing 80\% of the total galaxy mass.
The solid black line represents the broken power-law fit from \citet{RemyRuyer_2014}, where the slope at high metallicity reflects a constant DTM ratio.
The green line represents the linear regression fit from \citet{popping_2022}, derived from high-redshift galaxies in the range $0< z < 5$.
In \NC, the simulated relation well reproduces a near-linear trend above solar metallicity $\left(12 + \log{\rm (O/H)_\odot} \approx 8.69\right)$ where ${\rm Z_\odot \approx 0.0134}$ is adopted \citep{asplund_2009}.
In the low-metallicity regime, the simulated galaxies exhibit significantly lower DTG ratios compared to \citet{RemyRuyer_2014, popping_2022}, while comparable with \citet{deVis_2017}.
These trends are broadly consistent with previous results from cosmological simulations \citep[e.g.,][]{Popping_2017, hou2019, li2019, Dubois24, trayford_2025}.
Overall, this comparison suggests that the adopted dust model captures key aspects of dust evolution and is reasonable for further analysis.

\section{Methodology}
\label{sec:method}
\subsection{JWST Observation}
\label{sec:jwst}
To provide a reference for \NC, we utilize observational data that satisfy several essential criteria.
First, the dataset should include a large sample of galaxies with redshift estimates. Second, we prefer galaxies observed in a blank field to ensure that their morphologies are less affected by gravitational lensing effects. Lastly, the observation requires coverage in filters that probe the morphology of high-redshift galaxies.
We choose the JWST PRIMER-COSMOS East field (GO-1837, PI: Dunlop), which is part of the well-studied COSMOS survey \citep{2007ApJS..172....1S} and satisfies the selection criteria described above. 
Although \NC\ is a cluster simulation, it is hardly a cluster at $z=1$--3, which is the range investigated in this study. 
It is clearly a region where many galaxies formed early, which allow for a statistical analysis (despite its small volume) to compare with observational data. 
For a fairer comparison with \NC, we should ideally compare it to a set of observed cluster galaxies. However, identifying cluster galaxies at high redshifts requires spectroscopic redshifts, which are currently largely unavailable. 
Besides, we use the observed data only as a guideline, rather than as an aim of the simulation. Consequently, rather than pursuing a direct comparison with the simulation, we employ these galaxies as a morphological reference to interpret trends in LTG fractions.

All the JWST data are provided by the \textit{Dawn JWST archive} \citep{2023ApJ...947...20V} and the galaxy images are reduced using the \texttt{grizli} software \citep{brammer_2023_8370018}.
This field provides deep JWST imaging, making it an excellent choice for studying high-redshift galaxies.
We focus on galaxies within the redshift range $0.83 < z < 3.14$ and with stellar masses greater than $10^{10} M_{\odot}$. Given the wide redshift range of our sample, it is crucial to account for the dependence of morphology indices on rest-frame wavelength. Previous studies have shown that morphology indices become nearly independent of wavelength at $\lambda > 1000$ nm \citep{yaoyao_2023}. Within our redshift range, this corresponds to JWST's F277W and F356W filters. Therefore, we measure morphology indices in the F277W filter for lower redshifts $(z = 0.95 - 1.33)$ and in the F356W filter for higher redshifts $(z = 1.86 - 3.01)$. To ensure reliable measurements, we further restrict our sample to galaxies with  $18 < m < 30$, where $m$ denotes the apparent magnitude for a filter of our choice.  Additionally, we require a signal-to-noise ratio (SNR) greater than 10 in each filter to select galaxies with more reliable photometric redshift estimates. After applying these selection criteria, our final sample consists of 8,723 galaxies.

\subsection{NewCluster Mock Observation}
\begin{figure}[htbp]
    \centering
	\includegraphics[width=\columnwidth]{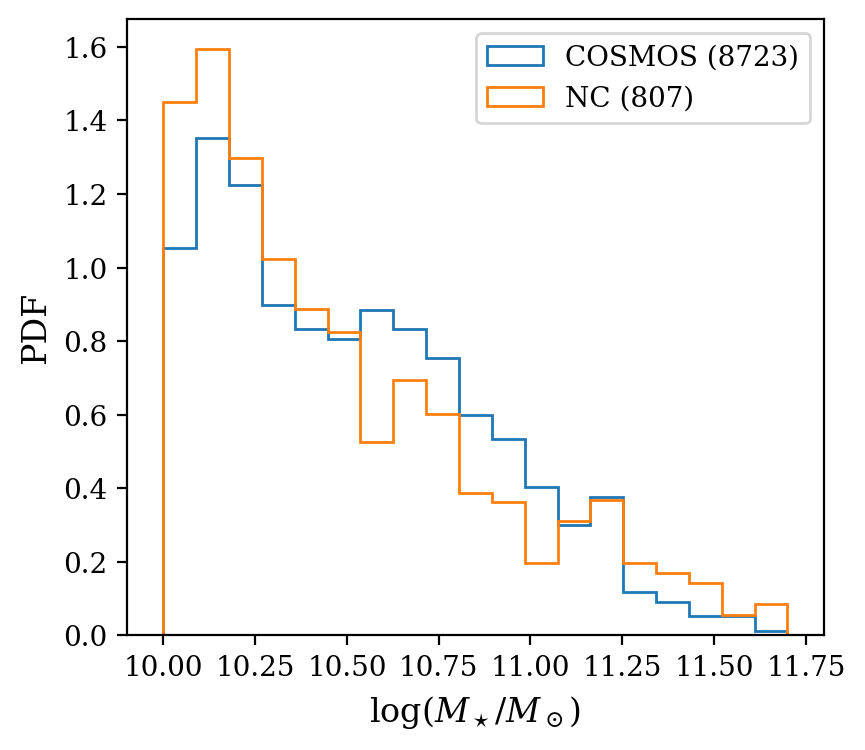}
   \caption{The stellar mass distribution of galaxies in \NC\ and JWST PRIMER-COSMOS East field. The number in the legend is the number of galaxies in each dataset.}
    \label{fig:mass_dist}
\end{figure}

To compare with the selected JWST observation and obtain realistic galaxy morphologies, we perform mock observations of \NC\ galaxies. Our sample consists of galaxies with stellar masses ranging from $10^{10}$ to $10^{12}\,\msol$ across five snapshots ($z = 0.95,1.33, 1.86, 2.44, {\rm and}\ 3.01$) of the \NC\ simulation. The stellar mass of each galaxy is defined as the total mass of member stellar particles identified by an AdaptaHOP galaxy-finder algorithm \citep{aubert2004adaptahop, tweed2009adaptaHOP}. 
In Figure~\ref{fig:mass_dist}, we present the stellar mass distribution of galaxies in the JWST observation and \NC. Their mass distributions are similar over the mass range of interest. As noted in Section \ref{sec:jwst}, because the JWST sample is used as a qualitative reference and not a statistically weighted control sample, we do not marginalize simulation data over stellar mass.

To perform the mock observation, we process the simulation data as follows:
\begin{itemize}
    \item Generating mock images of the \NC\ galaxies using the three-dimensional radiative transfer code \mbox{\sc \small SKIRT} \citep{SKIRT2015, camps2020SKIRT9} in JWST filters.
    \item Downscaling the simulated galaxy images from a pixel scale of $0.05\, \rm kpc/\rm pixel$ to match the observation pixel scale of $0\farcs04 ~\rm pixel^{-1}$. 
    \item Convolving the downscaled images with the mean point spread function modeled from the observation.
    \item Introducing characteristic JWST noise to the images.
\end{itemize}
The downscaling process simply involves using photometric galaxy redshifts under the assumption of a flat $\Lambda$CDM cosmology to convert from the original physical pixel scale to the necessary angular pixel scale. 

\subsubsection{SKIRT}

We generate the mock flux maps (i.e., mock images) of \NC\ galaxies using the \mbox{\sc \small SKIRT} radiative transfer code.
\mbox{\sc \small SKIRT} simulates the effect of dust on photons emitted from light sources, such as stellar particles, by accounting for absorption, emission, and scattering. We apply the Starburst99 stellar population model \citep{starburst99} with a Chabrier IMF \citep{chabrier2003IMF} to each stellar particle, to be consistent with \NC's setup. 

In \mbox{\sc \small SKIRT}, the number of photon packets, $N$, determines the statistical accuracy of the Monte Carlo radiative transfer simulation.
A convergence test up to $N = 10^9$ confirms that the resulting galaxy morphology shows no significant variation with increasing $N$. Based on this, we adopt $N = 2 \times 10^7$, which provides a good balance between noise suppression and computational efficiency.
The pixel scale for the mock images is set to 0.05 $\rm kpc / \rm pixel$ to ensure a detailed resolution. To improve statistical robustness, we generate mock images from 16 different viewing angles, covering four inclination angles and four azimuthal angles.

We employ two different dust models as input for \mbox{\sc \small SKIRT} to compare their effects on morphology. 
In the OTF model, we directly use the dust density distribution computed within \NC, as described in Section \ref{sec:nc_dust}.
In contrast, the fixed DTM model obtains the dust density distribution by scaling the metallicity distribution of \NC\ with a fixed DTM ratio. In this study, we adopt a fixed DTM value of 0.25, which corresponds to the average value measured for galaxies in \NC, as shown in Figure~\ref{fig:scaling_relation}. 
Additionally, we assume a maximum temperature of $30{,}000\, \rm{K}$ for the survival of dust species in the fixed DTM model.

Both the OTF and fixed DTM models adopt a two-component grain size population (small and large grains), and the small-to-large dust mass ratio is directly taken from the \NC\ outputs. This ratio reflects the physical processes that govern grain evolution, such as growth and shattering.
We input the small and large grain abundances into the post-processing radiative transfer code, SKIRT.
When doing so, we assume a log-normal grain size distribution with fixed minimum and maximum grain sizes of $1.0\times 10^{-7}\, \rm{cm}$ and $1.0\times 10^{-4}\, \rm{cm}$, respectively, for each grain type and (size) population. The peak sizes of the distributions are set to $5\, \rm{nm}$ for small grains and $0.1\, \rm{\mu m}$ for large grains.
For the SKIRT post-process, each distribution is discretized into 7 size bins per grain population. Although increasing the number of size bins could marginally improve accuracy, we find that seven bins are sufficient to capture the relevant galaxy morphology for our study.
Finally, both dust models adopt the grain optical properties from \citet{weingartnerDraine_2001}.

\begin{figure*}[htbp]
    \centering
	\includegraphics[width=\textwidth]{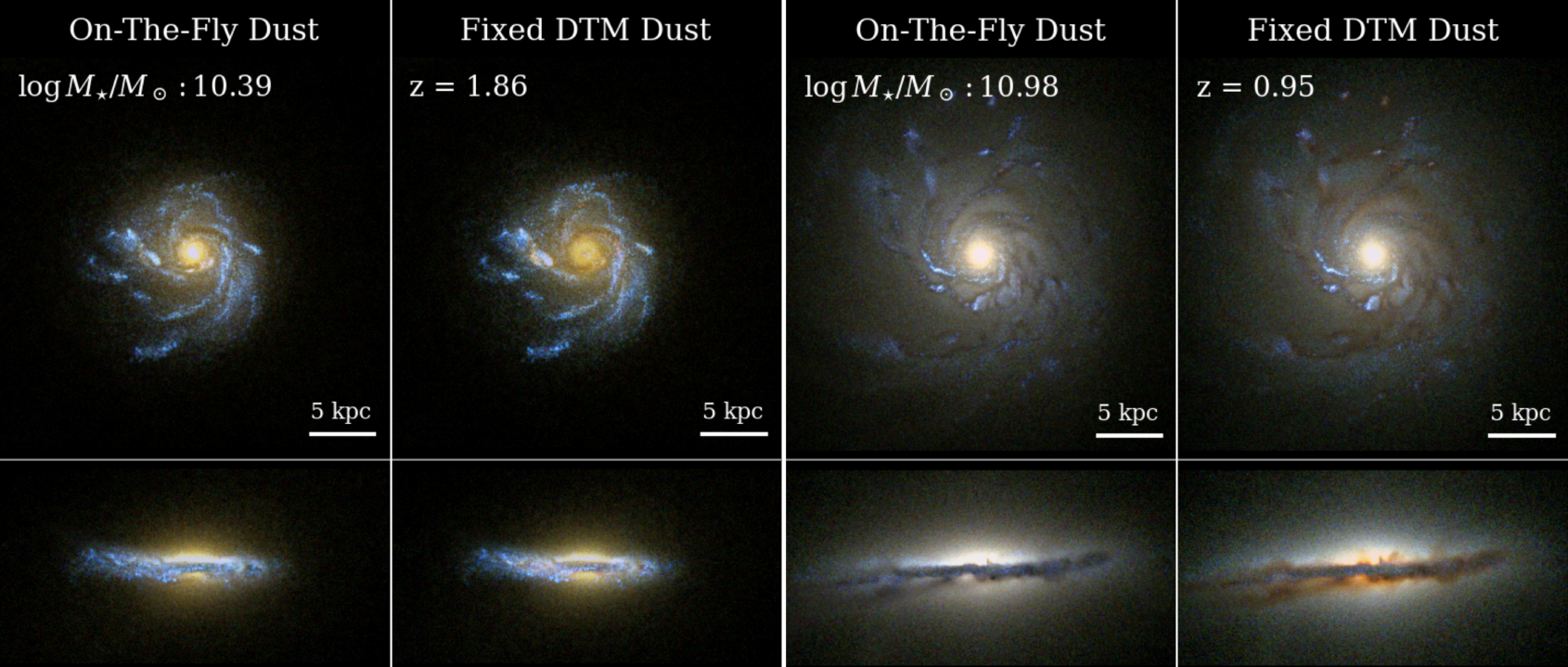}
    \caption{Mock images of two \NC\ simulation galaxies rendered with two dust models. The face-on and edge-on images are in the JWST filters. Red and blue colors correspond to JWST F200W and F070W filters. For green colors, JWST F090W and F150W filters are used. The first and third panels from the left display mock images generated using the OTF dust model, whereas the second and fourth panels correspond to the fixed DTM dust model.}
    \label{fig:skirt_img}
\end{figure*}

Using these dust properties, \mbox{\sc \small SKIRT} performs radiative transfer simulations to compute dust attenuation effects in the mock observations. In Figure \ref{fig:skirt_img}, we present the color images of two \NC\ galaxies in JWST filters, generated with \mbox{\sc \small SKIRT}. The comparison between the OTF and fixed DTM models reveals notable morphological differences. For the left sample galaxy, the OTF model produces a brighter central region compared to the fixed DTM model. For the right sample galaxy, the edge-on image generated with the fixed DTM model appears significantly redder, indicating stronger dust attenuation.

\subsubsection{Point spread function modeling}

A point spread function (PSF) is the response of a detector to a point source of light and represents the distortion introduced by the telescope. To model the PSF, we utilize stars in the observation field for a principal component analysis (PCA). PCA defines basis functions from the data itself rather than relying on predefined analytic functions to reconstruct the PSF.
Thus, we derive a PSF that fits the target observation, ensuring that the characteristics of the mock observations closely match those of the target observation. To achieve this, we perform a PCA on stars in the JWST PRIMER-COSMOS East field following the methodology of \cite{jee2007PASP..119.1403J} and \cite{finnerb-2023ApJ...953..102F}.

When introducing the PSF model, we convolve the downscaled image of the simulated galaxy with the mean PSF. This PSF is reconstructed using principal components that capture 90\% of the total variance, while the remaining components are discarded as they primarily contain noise rather than signal.

\subsubsection{Background noise}
In JWST observations, there are various sources of noise, such as Poisson noise and read noise. Over long exposures, the combination of these noise sources tends to approach a Gaussian distribution. 
While modeling the background noise with a simple Gaussian function can be sufficient in some cases, it is important to account for a realistic noise profile in morphological studies, as noise can significantly impact morphological measurements, especially for faint galaxies.
The background noise distribution in a mosaic image can deviate from following a Gaussian function for various reasons, with one significant cause being the drizzling procedure. Drizzling exposures onto a higher resolution grid leads to unphysical correlation between pixels, and accounting for this correlation is crucial when measuring galaxy morphology. To this end, we adopt the correlated background noise modeling method from \cite{article}, with the detailed implementation described in Appendix \ref{sec:correlated_noise}. 

\begin{figure*}
    \centering
	\includegraphics[width=0.9\textwidth]{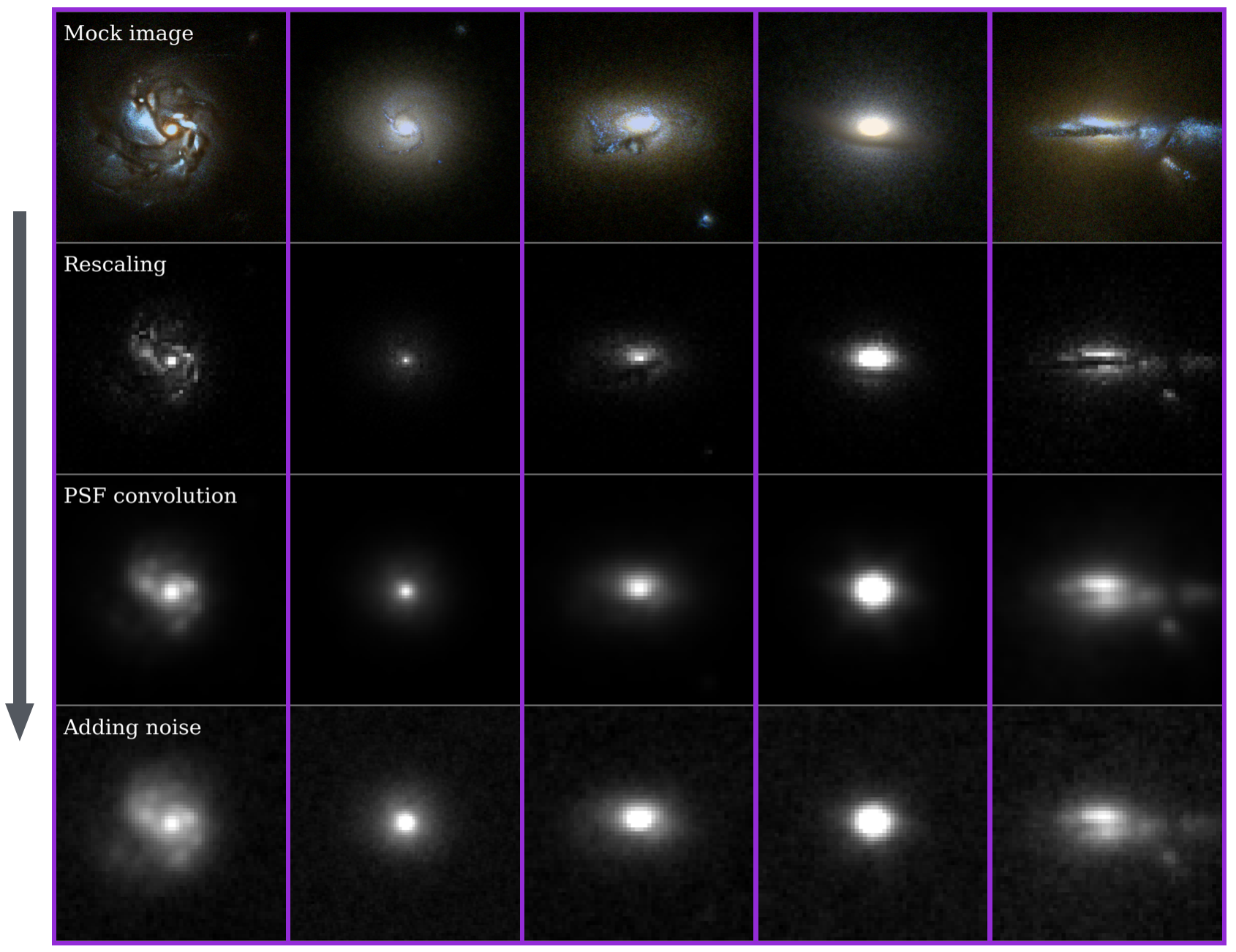}
    \caption{The overall mock observation process for five \NC\ galaxies. Each row represents the images at different stages of the mock observation process, illustrating how the images are transformed through each step. From top to bottom, each stage corresponds to \mbox{\sc \small SKIRT} simulation, rescaling, PSF convolution, and noise addition. The \mbox{\sc \small SKIRT} images in the first row are visualized in the same manner as those in Figure \ref{fig:skirt_img}, while the subsequent rows display images in the JWST F277W filter.}
    \label{fig:mock_process}
\end{figure*}

\begin{figure*}
    \centering
	\includegraphics[width=\textwidth]{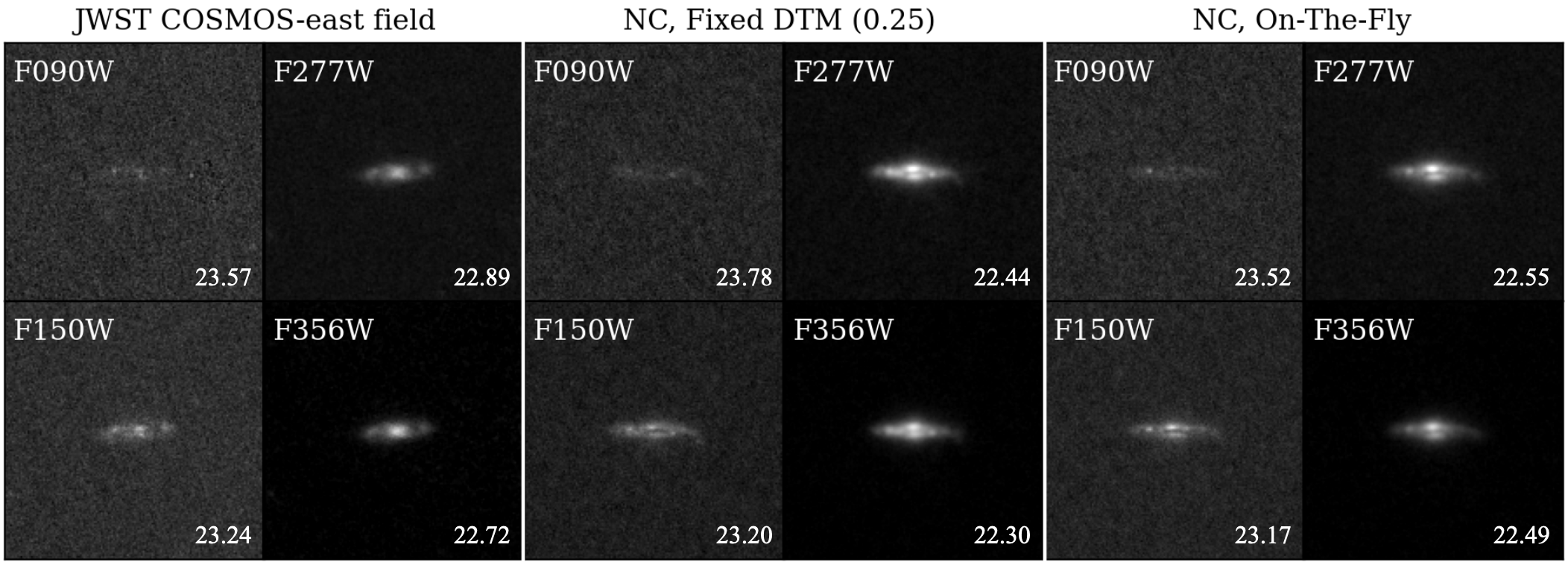}
    \caption{JWST COSMOS East galaxy images and the final mock-observed images from \NC. 
    The left four panels show a galaxy from the JWST COSMOS East field observed with four different filters. The middle four panels display a mock observed \NC\ galaxy utilizing a fixed DTM dust model. The right four panels show the same mock observed galaxy, instead utilizing the OTF dust model. The magnitudes are shown in the lower right corner of each panel.}
    \label{fig:mock_obs}
\end{figure*}

The overall mock observation process is summarized in Figure \ref{fig:mock_process}. Each row shows the images at successive stages of the process, illustrating how the galaxy images are transformed from top to bottom.
The final mock-observed images are presented in Figure \ref{fig:mock_obs}. In Figure \ref{fig:mock_obs}, the left four panels show an observed galaxy from the JWST COSMOS East field in four different filters. The middle four panels display the mock-observed images of a galaxy from \NC\ using the fixed DTM dust model, while the right four panels present the same but using the OTF dust model. The redshift of the observed JWST galaxy is 1.71, and the redshift of the \NC\ galaxy is 1.86. A comparison between the observed and model images shows that the noise levels and image quality are highly consistent, demonstrating the effectiveness of our mock observation procedure.

\subsection{Morphology indices}
\label{sec:morph_ind}
To quantitatively compare the morphologies of mock-observed galaxies from the two different dust models, we use the concentration index ($C$), asymmetry index ($A$) \citep{Conselice_2003}, the Gini coefficient ($G$) and the second-order moment of the brightest 20\% of each galaxy's flux ($M_{20}$) \citep{Lotz_2004}.

For each mock galaxy image, we create a segmented image using the \texttt{photutils} photometry package \citep{larry_bradley_2024_12585239}. Using these segmented images, we mask other sources and calculate the morphology indices.

The concentration index ($C$) is defined as 
\begin{equation}
    C \equiv 5\log_{10}\left(\frac{r_{80}}{r_{20}}\right),
\end{equation}
where $r_{80}$ and $r_{20}$ denote the circular radii enclosing 80\% and 20\% of the total galaxy flux, respectively. Galaxies with bright central regions and concentrated light distributions, such as ETGs, exhibit higher $C$ values. Among LTGs, those with prominent bulges also show elevated $C$ values.

The asymmetry index ($A$) is defined as 
\begin{equation}
    A \equiv \frac{\sum_{i,j} |I_{i,j} - I_{i,j}^{180}|}{\sum_{i,j} |I_{i,j}|},
\end{equation}
where $I_{i,j}$ is the original pixel flux, and $I_{i,j}^{180}$ is the flux at the same position in an image rotated by 180 degrees. The calculation includes only pixels within $r_{70}$ where $r_{70}$ denotes the circular radius enclosing 70\% of the total galaxy flux. Higher $A$ values indicate greater asymmetry, typically caused by features such as mergers or spiral arms in LTGs \citep{conselice_2000}.

The Gini coefficient ($G$) is calculated as
\begin{equation}
    G = \frac{1}{|\bar{I}|n(n-1)} \sum_{i=1}^{n} (2i - n - 1) |I_i|,
\end{equation}
where $I_i$ is the original pixel flux, and $n$ is the number of pixels. Same with $A$, the calculation of $G$ is restricted to pixels within $r_{70}$.
The Gini coefficient quantifies the inequality in a galaxy's flux distribution, with $G = 0$ indicating a perfectly uniform distribution. Similar to $C$, ETGs with bright centers or bulge-dominated LTGs exhibit higher $G$ values. However, $G$ is more sensitive to merger relic features than $C$ \citep{Lotz_2004}.

Finally, the second-order moment of the brightest 20\% of the galaxy's flux ($M_{20}$) is defined as
\begin{equation}
    M_{20} \equiv \log_{10} \left( \frac{\sum_{i} M_i}{M_{\rm tot}} \right),\ \text{until}\ \sum_{i} I_i < 0.2 I_{\rm tot},
\end{equation}
where the total second-order central moment $M_{\rm tot}$ is given by
\begin{equation}
    M_{\rm tot} = \sum_{i=1}^{n} M_i \equiv \sum_{i=1}^{n} I_i \left[ (x_i - x_c)^2 + (y_i - y_c)^2 \right],
\end{equation}
where $(x_i, y_i)$ denote the pixel coordinates and $(x_c, y_c)$ denote the coordinates of the galaxy center.
To compute $M_{20}$, pixels are sorted by flux, and $M_i$ is summed until the total flux reaches 20\% of $I_{\rm tot}$. $M_{20}$ generally decreases with increasing $C$. Like $G$, $M_{20}$ is sensitive to merger relics, as it is not constrained by a circular aperture centered on the galaxy \citep{Lotz_2004}.

There are lots of morphology classification schemes that use those morphology indices. However, high-redshift galaxies exhibit diverse morphologies, including merging systems, which makes some morphology classification schemes less effective. To address this, we adopt the \gm\ morphology from \citet{Lotz_2004}, for analyzing high-$z$ galaxies.  Unlike schemes that rely on parametric models, the \gm\ morphology is model-independent and robust against variations in resolution, making it particularly suitable for classifying galaxies in low-resolution high-redshift observations.

Given these advantages, we employ the \gm\ morphology to classify our sample. \gm\ morphology categorizes galaxies based on their positions in the two-dimensional \gm\ parameter plane. It distinguishes merging galaxies, early-type galaxies (E, S0, Sa Hubble types), and late-type galaxies (Sb, Sc, Irr Hubble types). 
Merging galaxies or those exhibiting merger relic features are identified by the criterion:
\begin{equation}
    \text{Merger}: G > -0.14M_{20} + 0.33.
\end{equation}
Non-merging galaxies, which fall below the merger classification line, are further classified as early-type or late-type galaxies. Late-type galaxies satisfy the condition:
\begin{equation}
    \text{Late\ type\ galaxy} : G < 0.14M_{20} + 0.80.
\end{equation}
The merger classification line and the morphology classification line are adopted from \citet{lotz2008gm20}. Galaxies classified as early-type by this scheme are bulge-dominated, while late-type galaxies tend to have disk-like or irregular morphologies.

\section{Results}
\label{sec:result}

\subsection{Morphological difference between dust models}
\label{sec:morph_diff}
\begin{figure}
	\includegraphics[width=\columnwidth]{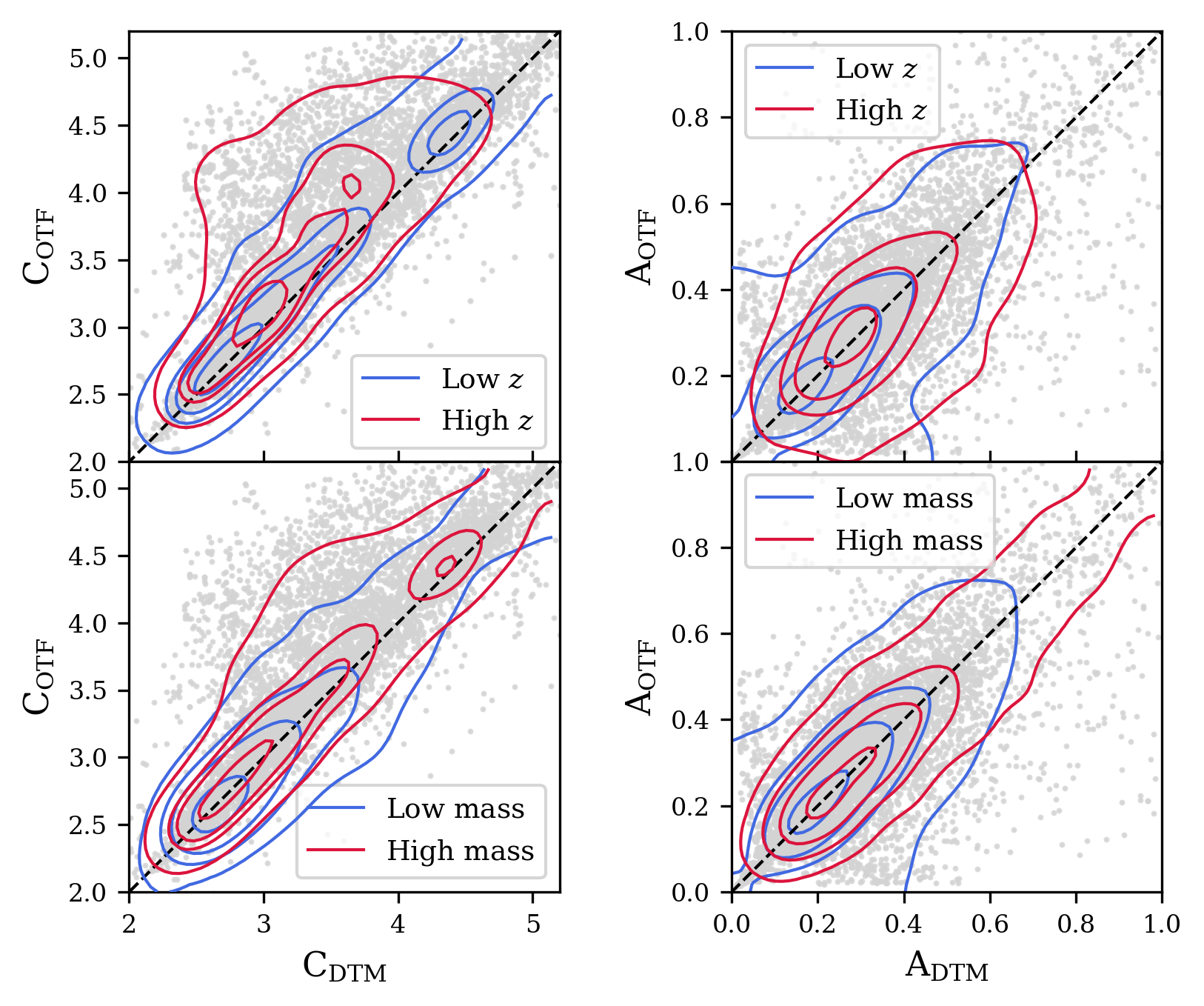}
    \caption{Comparison of the concentration ($C$) and asymmetry ($A$) measured in the JWST F277W filter for the same galaxies but with different dust models. The $x$-axis represents morphological indices derived from the fixed DTM model, while the $y$-axis represents those from the OTF model. 
    The top panels compare morphological indices across different redshift ranges, with red contours representing high redshift galaxies ($z = 1.86 - 3.01$) and blue contours representing low redshift galaxies ($z = 0.95 - 1.33$). The bottom panels categorize the comparisons based on stellar mass, where red contours indicate high mass galaxies ($\log{M_{\star}/M_{\odot}} > 10$) and blue contours indicate low mass galaxies ($\log{M_{\star}/M_{\odot}} < 10$). 
    Each contour level encloses 68\%, 90\%, 95\%, 99\% of the corresponding galaxy subset.
    The black dashed line ($y = x$) serves as a reference, indicating no offset between the two dust models. The gray points represent the entire galaxy sample.}
    \label{fig:morph_ind}
\end{figure}
Using the mock-observed \NC\ galaxies, we find that the most notable morphological difference between the OTF and fixed DTM models lies in the central brightness. As shown in Figure \ref{fig:skirt_img}, galaxies modeled with the OTF model exhibit significantly brighter centers and more prominent bulges compared to those modeled with the fixed DTM model. While the galaxies in Figure \ref{fig:skirt_img} were selected to highlight this effect, similar trends are observed across the sample.

To quantify these differences, Figure \ref{fig:morph_ind} presents a comparison of the concentration ($C$) and asymmetry ($A$) indices between the two dust models. The $x$-axis corresponds to the values from the fixed DTM model, while the $y$-axis represents those from the OTF model. The top panels show variations across redshifts, and the bottom panels categorize galaxies by stellar mass. The black dashed line indicates the one-to-one reference, allowing us to identify systematic differences between the two models. 

As shown in Figure \ref{fig:morph_ind}, the offset in $C$ appears in the high-redshift $(z=1.86 -3.01)$ and high-mass $(\log{M_{\star}/M_{\odot}} >10)$ samples, where OTF-modeled galaxies systematically deviate from the one-to-one line. This trend reflects the higher central brightness in the OTF model. 

In contrast, $A$ exhibits a symmetric distribution along the equality line. Since the differences between dust models mainly appear in central brightness, they do not affect $A$.

\subsection{\texorpdfstring{$G\text{\textendash} M_{20}$ distribution}{G-M20 distribution}}

\begin{figure*}
    \centering
	\includegraphics[width=0.9\textwidth]{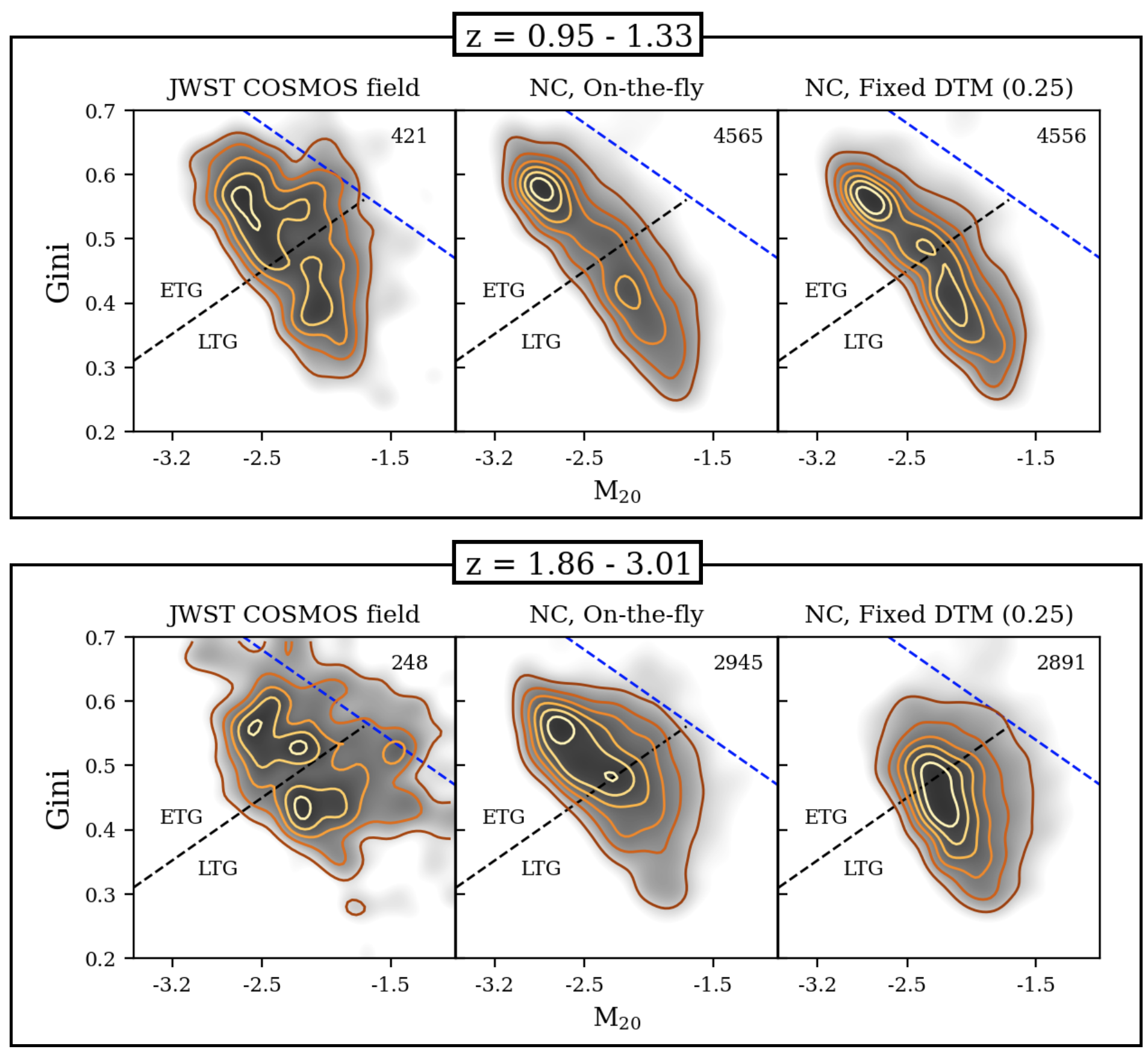}
    \caption{The distribution of galaxies in the \gm\ plane for JWST COSMOS East field observation and \NC\ galaxies with different dust models. The x-axis represents $M_{20}$, while the y-axis represents the Gini coefficient ($G$). The black dashed line separates ETGs and LTGs, while the blue dashed line denotes the merger classification threshold, as defined in Section~\ref{sec:morph_ind}. The top row corresponds to galaxies in the low-redshift range ($z = 0.95 - 1.33$), and the bottom row represents galaxies in the high-redshift range ($z = 1.86 - 3.01$). The leftmost panels show observational data from JWST, while the middle and rightmost panels show results from the \NC\ simulation using OTF and fixed DTM models, respectively. The contours indicate the number density of galaxies. All the galaxies have stellar mass higher than $10^{10}\, {M_\odot}$ in this figure.}
    \label{fig:gmstat}
\end{figure*}
To quantitatively compare the morphology of mock observed \NC\ galaxies with JWST COSMOS East field galaxies, we adopt the \gm\ morphology from \cite{Lotz_2004} as introduced in Section \ref{sec:morph_ind}. From this section onward, we use only the galaxies with masses greater than $10^{10}\, \rm{M_\odot}$, as 
the morphology classifications based on the \gm\ morphology and visual inspection are well aligned with each other above this mass cut. 

Figure \ref{fig:gmstat} presents the distribution of galaxies in the \gm\ plane for both observed JWST COSMOS East field galaxies and simulated galaxies from the \NC\ dataset using different dust models. The black dashed line separates ETGs and LTGs, while the blue dashed line represents the merger classification criterion, as defined in Section~\ref{sec:morph_ind}.
The figure is divided into two redshift ranges: the top row shows the galaxies at $z = 0.95 - 1.33$, and the bottom row for $z = 1.86 - 3.01$. The number density of galaxies is represented by the contour levels.

To quantify the similarity of two distributions, we measure the Jensen-Shannon distance (JSD), which quantifies the similarity between two probability distributions and ranges from 0 (identical) to 1 (maximally different). 
In Figure \ref{fig:gmstat}, the OTF dust model exhibits a similar distribution of galaxies on the \gm\ plane to the fixed DTM model in the low-redshift range ($z = 0.95$–$1.33$), as reflected in a relatively low JSD of 0.24. 
However, in the high-redshift range ($z = 1.86$–$3.01$), the JSD increases to 0.53, highlighting a more pronounced divergence in the \gm\ distributions between the two models. 
The OTF galaxies are more tightly clustered around the ETG region, whereas the DTM galaxies exhibit more extension towards the LTG region. 
This result arises from the morphological differences between the dust models, as discussed in Section \ref{sec:morph_diff}. A brighter galaxy center or a more prominent bulge-dominated feature leads to higher $G$ values and lower $M_{20}$ values. Consequently, galaxies utilizing the OTF model are more frequently classified as ETGs. Since morphological differences between the dust models become more significant at high redshifts, their impact on the \gm\ distribution is also more pronounced in this regime.

\NC\ simulates a dense cluster environment, whereas the JWST galaxy sample used for comparison primarily consists of galaxies in relatively low-density, field-like environments.
Due to this fundamental difference, the distributions of \NC\ and JWST galaxies may differ.
Nonetheless, when comparing the JWST observation with two different dust models from the \NC\ simulation, all galaxy samples exhibit a two-peak distribution with similar peak positions at low redshifts, with similar peak positions and similar JSD values of 0.54.
At high redshifts, however, the difference between the JWST and \NC\ distributions becomes more pronounced. Although both dust models yield high JSD values, the OTF model (0.83) shows a slightly lower divergence from the observed distribution than the fixed DTM model (0.85).
Meanwhile, both the observed and simulated galaxy samples display broader distributions at high redshift, particularly in regions associated with mergers, compared to their low-redshift counterparts.
This trend suggests a higher merger fraction and more diverse morphologies at higher redshifts, consistent with previous studies \citep{Conselice_2008, Bluck_2012, Ventou_2017}. Although both JWST and \NC\ show similar trends, \NC\ predicts fewer merger galaxies than observed. 
This discrepancy arises because, in simulations, galaxies are identified in 3D space, allowing clear separation even when galaxies are close in projection. In contrast, observations are limited to 2D projections, so galaxies that are physically distant can appear close along the line of sight, leading to a higher merger fraction \citep[see also][]{Rodriguez_2019}.

\subsection{Late-type galaxy fraction}
Using the \gm\ morphology shown in Figure \ref{fig:gmstat}, we define the LTG fraction as the ratio of galaxies classified in the LTG region to the total number of galaxies, excluding those in the merger region. 

\begin{figure}
	\includegraphics[width=\columnwidth]{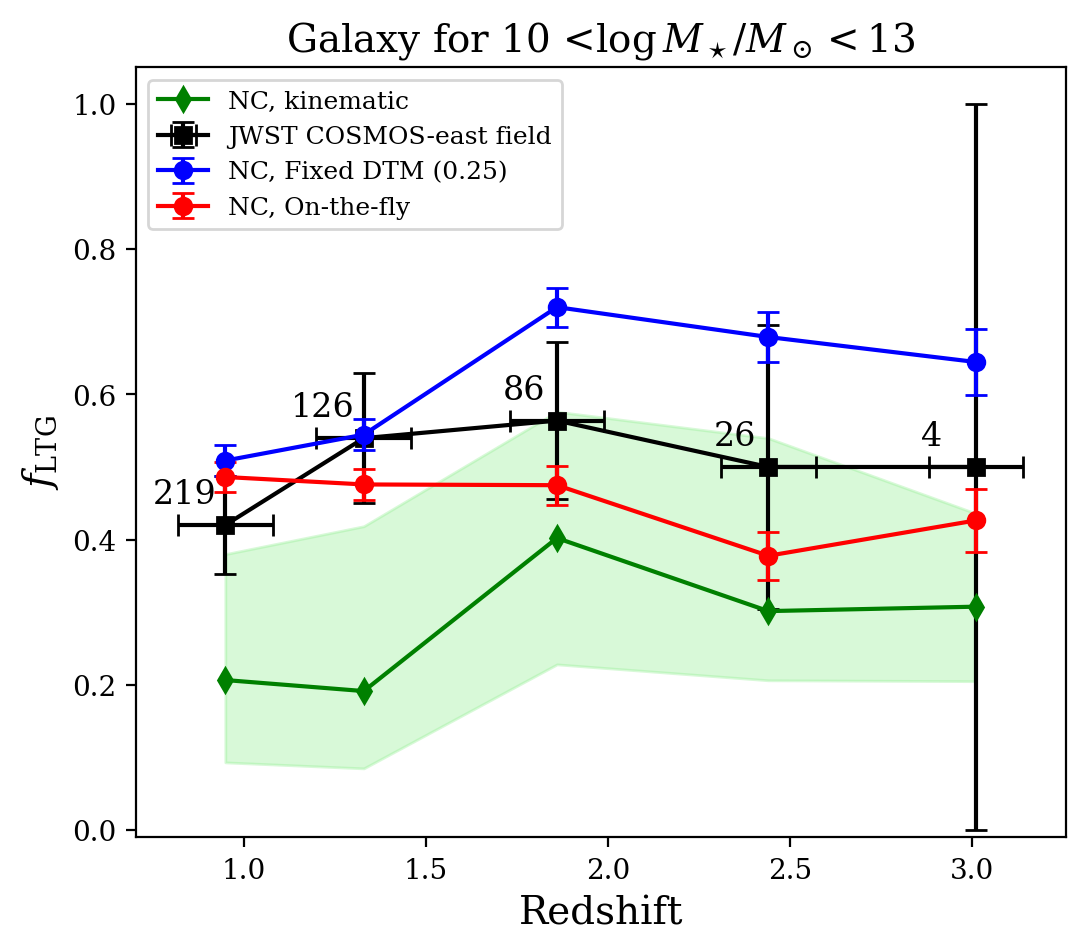}
    \caption{The fraction of late-type galaxies ($f_{\rm LTG}$) as a function of redshift for galaxies with stellar masses in the range $10 < \log{M_\star/M_\odot} < 13$. The black line with error bars represents the LTG fraction for JWST COSMOS East field galaxies, while the blue and red lines show LTG fractions for \NC\ galaxies using the fixed DTM and OTF dust models, respectively. The green line represents the LTG fraction based on kinematic classifications, with the shaded green region indicating variations due to different $V/\sigma$ criteria (see text). Poisson errors are shown as error bars, and the numbers above the black data points indicate the sample size for JWST COSMOS East field galaxies in each redshift bin. The shaded gray region marks the high-redshift range, where the small sample size of JWST observation increases uncertainty in this work.}
    \label{fig:fLTG_highm}
\end{figure}

Figure \ref{fig:fLTG_highm} shows the fraction of late-type galaxies ($f_{\rm LTG}$) as a function of redshift for galaxies with stellar masses in the range $10 < \log{M_\star/M_\odot} < 13$. 
The shaded gray region highlights the high-redshift range, where the sample size is significantly smaller, leading to increased uncertainty in the LTG fraction.
Additionally, the green line represents kinematic LTG fractions for \NC\ galaxies, determined based on the rotation-to-dispersion ratio of the stars, $V/\sigma$. 

As noted in Section~\ref{sec:jwst}, \NC\ simulates a dense cluster environment, whereas the JWST galaxies are located in relatively low-density, field-like environments. 
Given the well established correlation between galaxy morphology and environment—where denser regions tend to host lower LTG fractions \citep[e.g.,][]{Dressler_1980, Sazonova_2020}—a lower LTG fraction in \NC\ compared to the JWST observations is expected. This is consistent with the LTG fractions predicted by the OTF dust model, as well as with the kinematic LTG fractions presented in Figure~\ref{fig:fLTG_highm}. 
To ensure that our comparison is not biased by differences in stellar mass distributions, we first examine the distributions of the two samples and find an acceptable difference.
However, unlike the OTF model, the fixed DTM model predicts a significantly higher LTG fraction than the JWST observation at high redshifts $\left(z>1.5\right)$.
This discrepancy arises from the systematically fainter galaxy centers produced by the fixed DTM model, a trend consistently observed throughout our analysis, which was also visible in Figures \ref{fig:morph_ind} and \ref{fig:gmstat}. 

Kinematic morphology, like photometric morphology, provides crucial insights into galaxy formation and evolution. 
It is based on their internal kinematics, and in this study, we use the rotation-to-dispersion ratio, $V/\sigma$. The $V/\sigma$ values are computed in a cylindrical coordinate system aligned with the galactic rotation axis, where the mean rotational velocity is given by $V$ and the velocity dispersion is defined as $\sigma = \sqrt{\left(\sigma_{\rm r}^2 + \sigma_{\rm t}^2 + \sigma_{\rm z}^2\right)/3}$. Galaxies with $V/\sigma$ values exceeding a certain threshold are typically classified as kinematical LTGs.
Since the classification of kinematic LTGs is sensitive to the choice of $V/\sigma$ threshold, we present the kinematic LTG fraction of \NC\ galaxies using a threshold range of $V/\sigma = 1 \pm 0.1$ in this figure. 

As seen in Figure \ref{fig:fLTG_highm}, the kinematic LTG fraction appears lower than those derived from photometric classifications based on any dust model. This discrepancy may partly arise from the characteristics of the \gm\ morphology used for photometric classification. Although merger relic galaxies have been excluded, the \gm\ morphology can still classify some irregular galaxies as LTGs, potentially leading to an overestimation of the photometric LTG fraction compared to the kinematic classification.

More fundamentally, kinematic and photometric morphologies do not always align perfectly. In simulations, photometric morphology indicators tend to yield higher LTG fractions than kinematic indicators \citep{Thob_2019, Jang_2023}. Likewise, in observations, not all high-redshift disk galaxies are true rotational supported disks. They may exhibit flattened shapes due to anisotropy in stellar velocity dispersion rather than coherent rotation \citep{Wang_2024}. These discrepancies naturally lead to different evolutionary trends in the LTG fraction depending on the classification method, as seen in Figure \ref{fig:fLTG_highm}.

\begin{figure*}[htbp]
\centering
	\includegraphics[width=\textwidth]{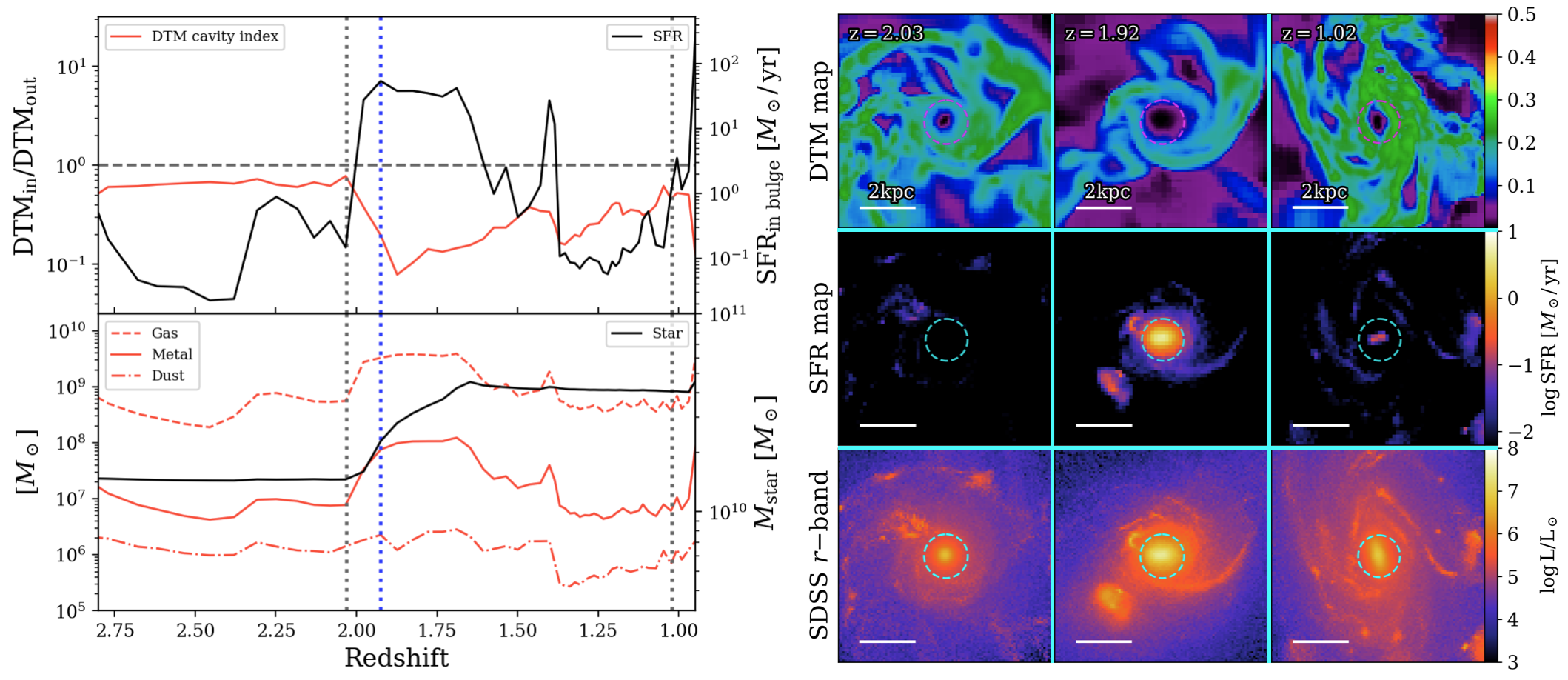}
    \caption{Formation and evolution of a central DTM cavity in a representative galaxy from the \NC\ simulation.
Left panels: Evolution of DTM cavity index and bulge properties. The top panel shows the evolution of the DTM cavity index (red; $DTM_{\rm in}/DTM_{\rm out}$) and the SFR within the bulge region (black). The bottom panel presents the evolution of gas-phase metal mass (solid red), dust mass (dot-dashed red), and gas mass (dashed red), while the stellar mass is plotted in black. Each vertical line indicates the epoch for the right panels.
Right panels: Projected maps of the same galaxy at three epochs ($z = 2.03$, 1.92, and 1.02; columns from left to right). From top to bottom: DTM, SFR, and SDSS $r$-band surface brightness. The dashed circle denotes the bulge radius. The central DTM cavity emerges shortly after a central starburst (marked by the blue vertical line in the left panels), and persists for several snapshots before being gradually filled in by dust growth.}
    \label{fig:dtmcavity}
\end{figure*}

\section{Discussion: DTM cavity formation}
\label{sec:discuss}

The most prominent morphological difference between the two dust models is that galaxies in the OTF model exhibit brighter galactic centers, leading to a lower LTG fraction. These brighter centers arise from a lower DTM ratio in the central region compared to the outer stellar regions. 
To quantitatively analyze this central dust depletion, we define ``DTM cavity index'', as the ratio $DTM_{\rm in}/DTM_{\rm out}$, where $DTM_{\rm in}$ and $DTM_{\rm out}$ are the mass-weighted average DTM ratios measured inside and outside the bulge radius, respectively. 
The DTM ratio is computed as the ratio of dust mass to total metal mass in both the gas and dust phases within each gas cell.
The ``bulge'' radius is determined at redshift $z=0.95$, as the location of the inflection point in the surface brightness profile measured in the SDSS $r$-band.

Figure \ref{fig:dtmcavity} illustrates the evolution of the DTM cavity in a sample galaxy from \NC.
In \NC, dust mass is modeled by four processes (Section~\ref{sec:nc_dust}): stellar ejecta, dust accretion, SN-driven destruction, and thermal sputtering. Therefore, we investigate the formation of DTM cavities in the context of these four channels.
The two panels on the left show the evolution of various dust-related quantities.
The top panel presents the DTM cavity index and the star formation rate (SFR) measured within the bulge region over a $10\, \rm{Myr}$ timescale. 
The bottom left panel tracks the evolution of gas-phase metal mass, dust mass, gas mass, and stellar mass. 
The nine panels on the right show the maps in DTM, SFR, and SDSS $r$-band flux at three epochs that are denoted in the left panels as vertical lines.

The top left panel shows that a strong central starburst occurs at $z \approx 2-1.7$, which coincides with a drop in DTM cavity index.
The DTM maps of the OTF models in the top-right panels show clearly the reduced DTM ratio at the center. 
This means a significantly lower dust extinction in the central region compared to the fixed DTM model.

The strong central starburst is likely associated with bulge formation.
During this short period, stellar and metal masses increase sharply, but dust mass hardly changes (bottom left panel). 
This is due to the destruction of dust by SN feedback, leading to a transient dip in the central DTM ratio (top right panels). 
The DTM cavity gradually fills in as dust growth becomes more efficient due to increased metallicity. 
These results indicate a feedback-driven mechanism for central dust depletion during bulge formation, followed by chemical evolution-driven dust accretion that replenishes the cavity over time.

\begin{figure}
	\includegraphics[width=\columnwidth]{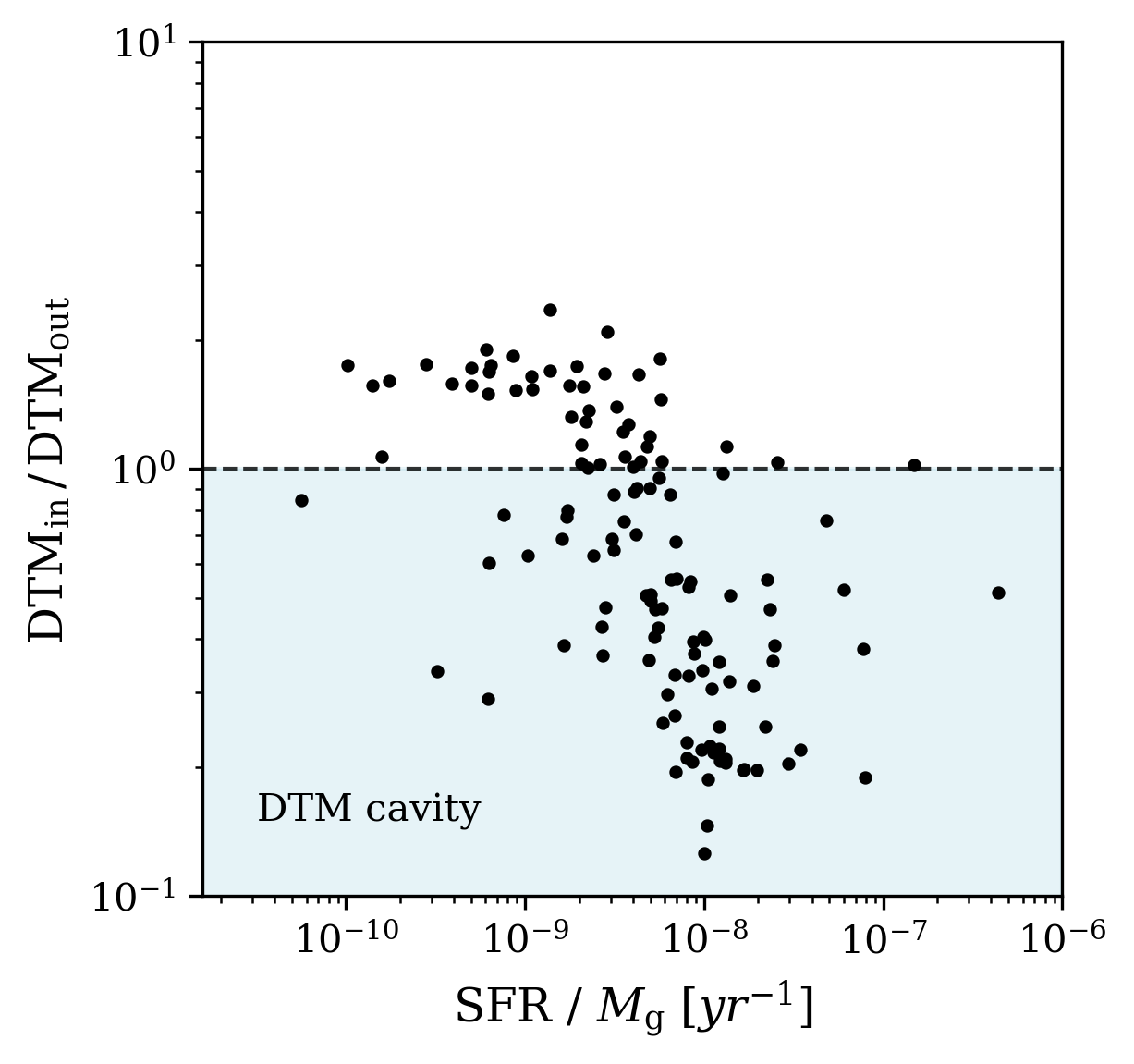}
    \caption{Relation between DTM cavity index and central SFR divided by gas mass for galaxies of stellar mass $M_\star > 10^{10}\,M_\odot$. The x-axis shows the SFR divided by the gas mass within the bulge, measured over a $10\, {\rm Myr}$ time window, while the y-axis indicates the DTM cavity index. Each point corresponds to a galaxy snapshot when the stellar mass within the bulge radius is growing most rapidly. The shaded region indicates DTM cavities.}
    \label{fig:bulgeform}
\end{figure}

This phenomenon is common among massive bulge-hosting galaxies in \NC. 
Figure \ref{fig:bulgeform} shows the relation between the DTM cavity index and central SFR divided by gas mass for galaxies with $M_\star > 10^{10}\,M_\odot$ at redshift $z = 0.95$. Each data point represents a snapshot of a galaxy during the phase of its most rapid stellar mass growth within the bulge radius, capturing an active phase of bulge formation.
A bulge with a stronger (gas-specific) SFR shows a more pronounced DTM cavity.
Observational studies have similarly proposed that starburst-driven feedback can drive central dust depletion \citep{Mattsson_2012}, and some bulge-dominated S0–Sa galaxies show central depressions in their dust radial profiles \citep{Mateos_2009}, lending further support to our scenario.

\begin{figure}
	\includegraphics[width=\columnwidth]{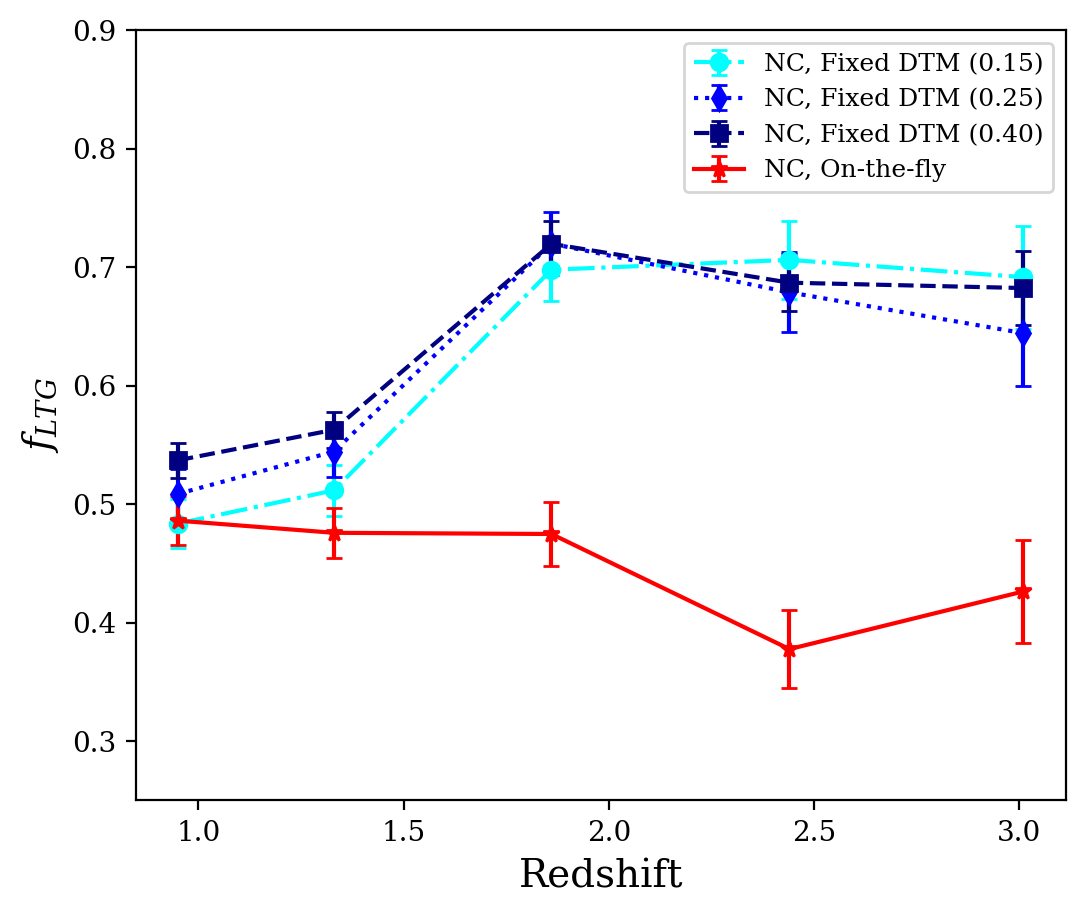}
    \caption{Similar format to Fig. \ref{fig:fLTG_highm}. Different lines represent the LTG fraction from \NC\ using various dust models: Fixed DTM with values of 0.15 (cyan), 0.25 (blue), 0.40 (navy), and the OTF model (red). Poisson errors, calculated based on the number of sample galaxies in each redshift bin, are shown as error bars. This comparison highlights the impact of different dust models on the predicted morphological evolution of galaxies.}
    \label{fig:DTMs}
\end{figure}

By construction, fixed-DTM models cannot reproduce DTM cavities.
As shown in Figure~\ref{fig:dtmcavity} (bottom left panel), metal mass alone does not lead to dust mass correctly without taking dust destruction into account. 
For the same reason, even trying different values of {\em fixed} DTM ratio does not help much, as illustrated in Figure \ref{fig:DTMs}.
The use of different values of fixed DTM (0.15,  0.25, and 0.40) exhibits similar trends of LTG fractions.

Altogether, these results indicate that bulge-driven starbursts not only enrich the central ISM with metals but also preferentially deplete dust through feedback, leading to the formation of distinct DTM cavities. These DTM cavities gradually disappear due to dust accretion, therefore, they are more frequently expected in the early universe. 
Considering the complexity of dust evolution, a proper modeling of dust is necessary to determine the morphology of galaxies from their images, especially at high redshifts where star formation rates were high.

\section{Conclusion}
\label{sec:con}

This study explores the impact of different dust models on galaxy morphology using the \NC\ cosmological hydrodynamics simulation, which incorporates an advanced on-the-fly dust model. By comparing the OTF and fixed DTM ratio models, we investigate how dust influences the observed structure of galaxies, particularly at high redshifts. Mock observations were generated using \mbox{\sc \small SKIRT} radiative transfer simulations, and morphology indices—including concentration ($C$), asymmetry ($A$), Gini coefficient ($G$), and $M_{20}$—were analyzed to classify galaxies and compare them with JWST observations.

The key findings are summarized as follows.
\begin{itemize}
    \item The fixed DTM model leads to fainter galactic centers and less pronounced bulges compared to those using the OTF dust model, especially in high-mass galaxies at high redshifts.
    \item Late-type galaxy fractions defined by the \gm\ test differ significantly between the two models, with the fixed DTM model predicting higher LTG fractions as a result of producing galaxies with fainter central regions. This discrepancy between dust models becomes more pronounced at $z > 1.5$.
    \item The enhanced central brightness in OTF galaxies is primarily driven by the presence of DTM cavities, which form due to central starburst and efficient SN-driven dust destruction. The fixed DTM model cannot reproduce these cavities, leading to systematic differences in galaxy concentration between the two dust models.
\end{itemize}

DTM cavities are expected to be more frequent in the early universe, where bulge formation was more active, but today's galaxies may also show it if they experience unusually enhanced star formation at the center.

Future work will focus on improving dust subgrid models, such as incorporating Coulomb enhancement factors in the dust accretion process. We are conducting a spin-off simulation using revised dust prescriptions; we plan to investigate their impacts on the evolution of dust, especially in the early universe. 
We also find that DTM cavities are present in this spin-off simulation, suggesting that the DTM cavity observed in \NC\ is unlikely to be an artifact or a feature specific to the particular dust model employed.
In addition, although this paper focuses on the formation of DTM cavities, our forthcoming study will explain why grain growth is not efficient immediately after starbursts and how DTM cavities evolve.

Our results indicate that the OTF model provides a more realistic description of galaxy morphology, showing better agreement with the observed trends in the \gm\ plane and LTG fractions.
These findings underscore the necessity of on-the-fly dust evolution models for accurately simulating galaxy structure and evolution, particularly in the early Universe where the dust properties were substantially different from the empirical data measured in the local Universe.

\section*{Acknowledgements}
We thank the referee for the constructive and insightful comments, which significantly helped us improve the clarity and depth of our work. In particular, Appendix B has been added in response to the referee's suggestion.
Some of the data products presented herein were retrieved from the Dawn JWST Archive (DJA). DJA is an initiative of the Cosmic Dawn Center (DAWN), which is funded by the Danish National Research Foundation under grant DNRF140.
S.K.Y. acknowledges support from the Korean National Research Foundation (RS-2025-00514475; RS-2022-NR070872).
J.R. was supported by the KASI-Yonsei Postdoctoral Fellowship and was supported by the Korea Astronomy and Space Science Institute under the R\&D program (Project No. 2025-1-831-02), supervised by the Korea AeroSpace Administration. This work was partially supported by the Institut de Physique des deux infinis of Sorbonne Université and by the ANR grant ANR-19-CE31-0017 of the French Agence Nationale de la Recherche.
This work was granted access to the HPC resources of KISTI under the allocations KSC-2021-CRE-0486, KSC-2022-CRE-0088, KSC-2022-CRE-0344, KSC-2022-CRE-0409, KSC-2023-CRE-0343, KSC-2024-CHA-0009, and KSC-2025-CRE-0031 and of GENCI under the allocation A0150414625 and A0180416216.
The large data transfer was supported by KREONET, which is managed and operated by KISTI. 

\bibliography{reference}{}

\begin{thebibliography}{}
\expandafter\ifx\csname natexlab\endcsname\relax\def\natexlab#1{#1}\fi
\providecommand{\url}[1]{\href{#1}{#1}}
\providecommand{\dodoi}[1]{doi:~\href{http://doi.org/#1}{\nolinkurl{#1}}}
\providecommand{\doeprint}[1]{\href{http://ascl.net/#1}{\nolinkurl{http://ascl.net/#1}}}
\providecommand{\doarXiv}[1]{\href{https://arxiv.org/abs/#1}{\nolinkurl{https://arxiv.org/abs/#1}}}

\bibitem[{{Agertz} {et~al.}(2011){Agertz}, {Teyssier}, \& {Moore}}]{Agertz_2011}
{Agertz}, O., {Teyssier}, R., \& {Moore}, B. 2011, \mnras, 410, 1391, \dodoi{10.1111/j.1365-2966.2010.17530.x}

\bibitem[{{Algera} {et~al.}(2025){Algera}, {Rowland}, {Stefanon}, {Palla}, {Sommovigo}, {Inami}, {Bouwens}, {Aravena}, {Bowler}, {Dayal}, {De Looze}, {Ferrara}, {Fisher}, {Graziani}, {Gulis}, {Heintz}, {Hodge}, {van Leeuwen}, {Pallottini}, {Phillips}, {Schouws}, {Smit}, {Stark}, \& {van der Werf}}]{Algera_2025}
{Algera}, H., {Rowland}, L., {Stefanon}, M., {et~al.} 2025, arXiv e-prints, arXiv:2501.10508, \dodoi{10.48550/arXiv.2501.10508}

\bibitem[{{Aoyama} {et~al.}(2020){Aoyama}, {Hirashita}, \& {Nagamine}}]{aoyama2020}
{Aoyama}, S., {Hirashita}, H., \& {Nagamine}, K. 2020, \mnras, 491, 3844, \dodoi{10.1093/mnras/stz3253}

\bibitem[{{Aoyama} {et~al.}(2017){Aoyama}, {Hou}, {Shimizu}, {Hirashita}, {Todoroki}, {Choi}, \& {Nagamine}}]{Aoyama_2017}
{Aoyama}, S., {Hou}, K.-C., {Shimizu}, I., {et~al.} 2017, \mnras, 466, 105, \dodoi{10.1093/mnras/stw3061}

\bibitem[{{Asplund} {et~al.}(2009){Asplund}, {Grevesse}, {Sauval}, \& {Scott}}]{asplund_2009}
{Asplund}, M., {Grevesse}, N., {Sauval}, A.~J., \& {Scott}, P. 2009, \araa, 47, 481, \dodoi{10.1146/annurev.astro.46.060407.145222}

\bibitem[{{Aubert} {et~al.}(2004){Aubert}, {Pichon}, \& {Colombi}}]{aubert2004adaptahop}
{Aubert}, D., {Pichon}, C., \& {Colombi}, S. 2004, \mnras, 352, 376, \dodoi{10.1111/j.1365-2966.2004.07883.x}

\bibitem[{{Baes} \& {Camps}(2015)}]{SKIRT2015}
{Baes}, M., \& {Camps}, P. 2015, Astronomy and Computing, 12, 33, \dodoi{10.1016/j.ascom.2015.05.006}

\bibitem[{{Bekki}(2013)}]{Bekki_2013}
{Bekki}, K. 2013, \mnras, 432, 2298, \dodoi{10.1093/mnras/stt589}

\bibitem[{{Bertin} \& {Arnouts}(1996)}]{1996A&AS..117..393B}
{Bertin}, E., \& {Arnouts}, S. 1996, \aaps, 117, 393, \dodoi{10.1051/aas:1996164}

\bibitem[{{Bluck} {et~al.}(2012){Bluck}, {Conselice}, {Buitrago}, {Gr{\"u}tzbauch}, {Hoyos}, {Mortlock}, \& {Bauer}}]{Bluck_2012}
{Bluck}, A. F.~L., {Conselice}, C.~J., {Buitrago}, F., {et~al.} 2012, \apj, 747, 34, \dodoi{10.1088/0004-637X/747/1/34}

\bibitem[{Bradley {et~al.}(2024)Bradley, Sip{\H o}cz, Robitaille, Tollerud, Vin{\'{\i}}cius, Deil, Barbary, Wilson, Busko, Donath, G{\"u}nther, Cara, Lim, Me{\ss}linger, Burnett, Conseil, Droettboom, Bostroem, Bray, Bratholm, Jamieson, Ginsburg, Barentsen, Craig, Pascual, Rathi, Perrin, Morris, \& Perren}]{larry_bradley_2024_12585239}
Bradley, L., Sip{\H o}cz, B., Robitaille, T., {et~al.} 2024, astropy/photutils: 1.13.0, 1.13.0,  Zenodo, \dodoi{10.5281/zenodo.12585239}

\bibitem[{Brammer(2023)}]{brammer_2023_8370018}
Brammer, G. 2023, grizli, 1.9.11,  Zenodo, \dodoi{10.5281/zenodo.8370018}

\bibitem[{{Camps} \& {Baes}(2020)}]{camps2020SKIRT9}
{Camps}, P., \& {Baes}, M. 2020, Astronomy and Computing, 31, 100381, \dodoi{10.1016/j.ascom.2020.100381}

\bibitem[{{Camps} {et~al.}(2016){Camps}, {Trayford}, {Baes}, {Theuns}, {Schaller}, \& {Schaye}}]{Camps_2016}
{Camps}, P., {Trayford}, J.~W., {Baes}, M., {et~al.} 2016, \mnras, 462, 1057, \dodoi{10.1093/mnras/stw1735}

\bibitem[{{Chabrier}(2003)}]{chabrier2003IMF}
{Chabrier}, G. 2003, \pasp, 115, 763, \dodoi{10.1086/376392}

\bibitem[{{Choban} {et~al.}(2024){Choban}, {Kere{\v{s}}}, {Sandstrom}, {Hopkins}, {Hayward}, \& {Faucher-Gigu{\`e}re}}]{choban2024a}
{Choban}, C.~R., {Kere{\v{s}}}, D., {Sandstrom}, K.~M., {et~al.} 2024, \mnras, 529, 2356, \dodoi{10.1093/mnras/stae716}

\bibitem[{{Conselice}(2003)}]{Conselice_2003}
{Conselice}, C.~J. 2003, \apjs, 147, 1, \dodoi{10.1086/375001}

\bibitem[{{Conselice} {et~al.}(2000){Conselice}, {Bershady}, \& {Jangren}}]{conselice_2000}
{Conselice}, C.~J., {Bershady}, M.~A., \& {Jangren}, A. 2000, \apj, 529, 886, \dodoi{10.1086/308300}

\bibitem[{{Conselice} {et~al.}(2008){Conselice}, {Rajgor}, \& {Myers}}]{Conselice_2008}
{Conselice}, C.~J., {Rajgor}, S., \& {Myers}, R. 2008, \mnras, 386, 909, \dodoi{10.1111/j.1365-2966.2008.13069.x}

\bibitem[{{Davis} {et~al.}(1985){Davis}, {Efstathiou}, {Frenk}, \& {White}}]{Davis_1985}
{Davis}, M., {Efstathiou}, G., {Frenk}, C.~S., \& {White}, S.~D.~M. 1985, \apj, 292, 371, \dodoi{10.1086/163168}

\bibitem[{{de Vaucouleurs}(1959)}]{1959HDP....53..275D}
{de Vaucouleurs}, G. 1959, Handbuch der Physik, 53, 275, \dodoi{10.1007/978-3-642-45932-0_7}

\bibitem[{{De Vis} {et~al.}(2017){De Vis}, {Gomez}, {Schofield}, {Maddox}, {Dunne}, {Baes}, {Cigan}, {Clark}, {Gomez}, {Lara-L{\'o}pez}, \& {Owers}}]{deVis_2017}
{De Vis}, P., {Gomez}, H.~L., {Schofield}, S.~P., {et~al.} 2017, \mnras, 471, 1743, \dodoi{10.1093/mnras/stx981}

\bibitem[{{De Vis} {et~al.}(2019){De Vis}, {Jones}, {Viaene}, {Casasola}, {Clark}, {Baes}, {Bianchi}, {Cassara}, {Davies}, {De Looze}, {Galametz}, {Galliano}, {Lianou}, {Madden}, {Manilla-Robles}, {Mosenkov}, {Nersesian}, {Roychowdhury}, {Xilouris}, \& {Ysard}}]{Devis_2019}
{De Vis}, P., {Jones}, A., {Viaene}, S., {et~al.} 2019, \aap, 623, A5, \dodoi{10.1051/0004-6361/201834444}

\bibitem[{{Dressler}(1980)}]{Dressler_1980}
{Dressler}, A. 1980, \apj, 236, 351, \dodoi{10.1086/157753}

\bibitem[{{Dressler} {et~al.}(1994){Dressler}, {Oemler}, {Butcher}, \& {Gunn}}]{Dressler_1994}
{Dressler}, A., {Oemler}, Jr., A., {Butcher}, H.~R., \& {Gunn}, J.~E. 1994, \apj, 430, 107, \dodoi{10.1086/174386}

\bibitem[{{Dubois} {et~al.}(2012){Dubois}, {Devriendt}, {Slyz}, \& {Teyssier}}]{dubois2012_AGN}
{Dubois}, Y., {Devriendt}, J., {Slyz}, A., \& {Teyssier}, R. 2012, \mnras, 420, 2662, \dodoi{10.1111/j.1365-2966.2011.20236.x}

\bibitem[{{Dubois} {et~al.}(2024){Dubois}, {Rodr{\'\i}guez Montero}, {Guerra}, {Trebitsch}, {Han}, {Beckmann}, {Yi}, {Lewis}, \& {Jang}}]{Dubois24}
{Dubois}, Y., {Rodr{\'\i}guez Montero}, F., {Guerra}, C., {et~al.} 2024, \aap, 687, A240, \dodoi{10.1051/0004-6361/202449784}

\bibitem[{{Dwek}(1998)}]{Dwek_1998}
{Dwek}, E. 1998, \apj, 501, 643, \dodoi{10.1086/305829}

\bibitem[{{Fall} \& {Efstathiou}(1980)}]{Fall_efstathiou_1980}
{Fall}, S.~M., \& {Efstathiou}, G. 1980, \mnras, 193, 189, \dodoi{10.1093/mnras/193.2.189}

\bibitem[{{Federrath} \& {Klessen}(2012)}]{federrath_2012}
{Federrath}, C., \& {Klessen}, R.~S. 2012, \apj, 761, 156, \dodoi{10.1088/0004-637X/761/2/156}

\bibitem[{{Ferreira} {et~al.}(2022){Ferreira}, {Adams}, {Conselice}, {Sazonova}, {Austin}, {Caruana}, {Ferrari}, {Verma}, {Trussler}, {Broadhurst}, {Diego}, {Frye}, {Pascale}, {Wilkins}, {Windhorst}, \& {Zitrin}}]{Ferreira_2022}
{Ferreira}, L., {Adams}, N., {Conselice}, C.~J., {et~al.} 2022, \apjl, 938, L2, \dodoi{10.3847/2041-8213/ac947c}

\bibitem[{{Finner} {et~al.}(2023){Finner}, {Faisst}, {Chary}, \& {Jee}}]{finnerb-2023ApJ...953..102F}
{Finner}, K., {Faisst}, A., {Chary}, R.-R., \& {Jee}, M.~J. 2023, \apj, 953, 102, \dodoi{10.3847/1538-4357/ace1e6}

\bibitem[{{Granato} {et~al.}(2021){Granato}, {Ragone-Figueroa}, {Taverna}, {Silva}, {Valentini}, {Borgani}, {Monaco}, {Murante}, \& {Tornatore}}]{Granato2021}
{Granato}, G.~L., {Ragone-Figueroa}, C., {Taverna}, A., {et~al.} 2021, \mnras, 503, 511, \dodoi{10.1093/mnras/stab362}

\bibitem[{{Haardt} \& {Madau}(1996)}]{haardt_madau1996}
{Haardt}, F., \& {Madau}, P. 1996, \apj, 461, 20, \dodoi{10.1086/177035}

\bibitem[{{Han} {et~al.}(2025{\natexlab{a}}){Han}, {Dubois}, {Lee}, {Kim}, {Cadiou}, \& {Yi}}]{ramsesyomp}
{Han}, S., {Dubois}, Y., {Lee}, J., {et~al.} 2025{\natexlab{a}}, \apj, 978, 96, \dodoi{10.3847/1538-4357/ad98f4}

\bibitem[{{Han} {et~al.}(2025{\natexlab{b}}){Han}, {Yi}, {Dubois}, {Rhee}, {Jeon}, {Jang}, {Byun}, {Cadiou}, {Kim}, {Kimm}, \& {Pichon}}]{nc_arXiv_2025}
{Han}, S., {Yi}, S.~K., {Dubois}, Y., {et~al.} 2025{\natexlab{b}}, arXiv e-prints, arXiv:2507.06301.
\newblock \doarXiv{2507.06301}

\bibitem[{{Hayward} {et~al.}(2014){Hayward}, {Lanz}, {Ashby}, {Fazio}, {Hernquist}, {Mart{\'\i}nez-Galarza}, {Noeske}, {Smith}, {Wuyts}, \& {Zezas}}]{Hayward_2014}
{Hayward}, C.~C., {Lanz}, L., {Ashby}, M. L.~N., {et~al.} 2014, \mnras, 445, 1598, \dodoi{10.1093/mnras/stu1843}

\bibitem[{{Hirashita}(2015)}]{Hirashita15}
{Hirashita}, H. 2015, \mnras, 447, 2937, \dodoi{10.1093/mnras/stu2617}

\bibitem[{{Hirashita} \& {Kuo}(2011)}]{hirashita_2011}
{Hirashita}, H., \& {Kuo}, T.-M. 2011, \mnras, 416, 1340, \dodoi{10.1111/j.1365-2966.2011.19131.x}

\bibitem[{{Hou} {et~al.}(2019){Hou}, {Aoyama}, {Hirashita}, {Nagamine}, \& {Shimizu}}]{hou2019}
{Hou}, K.-C., {Aoyama}, S., {Hirashita}, H., {Nagamine}, K., \& {Shimizu}, I. 2019, \mnras, 485, 1727, \dodoi{10.1093/mnras/stz121}

\bibitem[{{Hubble}(1926)}]{Hubble_1926}
{Hubble}, E.~P. 1926, \apj, 64, 321, \dodoi{10.1086/143018}

\bibitem[{{Inoue}(2003)}]{inoue_2003}
{Inoue}, A.~K. 2003, \pasj, 55, 901, \dodoi{10.1093/pasj/55.5.901}

\bibitem[{{Iwamoto} {et~al.}(1999){Iwamoto}, {Brachwitz}, {Nomoto}, {Kishimoto}, {Umeda}, {Hix}, \& {Thielemann}}]{iwamoto1999}
{Iwamoto}, K., {Brachwitz}, F., {Nomoto}, K., {et~al.} 1999, \apjs, 125, 439, \dodoi{10.1086/313278}

\bibitem[{{Jang} {et~al.}(2023){Jang}, {Yi}, {Dubois}, {Rhee}, {Pichon}, {Kimm}, {Devriendt}, {Volonteri}, {Kaviraj}, {Peirani}, {Oh}, \& {Croom}}]{Jang_2023}
{Jang}, J.~K., {Yi}, S.~K., {Dubois}, Y., {et~al.} 2023, \apj, 950, 4, \dodoi{10.3847/1538-4357/accd68}

\bibitem[{{Jee} {et~al.}(2007){Jee}, {Blakeslee}, {Sirianni}, {Martel}, {White}, \& {Ford}}]{jee2007PASP..119.1403J}
{Jee}, M.~J., {Blakeslee}, J.~P., {Sirianni}, M., {et~al.} 2007, \pasp, 119, 1403, \dodoi{10.1086/524849}

\bibitem[{{Kassin} {et~al.}(2012){Kassin}, {Weiner}, {Faber}, {Gardner}, {Willmer}, {Coil}, {Cooper}, {Devriendt}, {Dutton}, {Guhathakurta}, {Koo}, {Metevier}, {Noeske}, \& {Primack}}]{Kassin_2012}
{Kassin}, S.~A., {Weiner}, B.~J., {Faber}, S.~M., {et~al.} 2012, \apj, 758, 106, \dodoi{10.1088/0004-637X/758/2/106}

\bibitem[{{Kauffmann} {et~al.}(1993){Kauffmann}, {White}, \& {Guiderdoni}}]{Kauffmann_1993}
{Kauffmann}, G., {White}, S.~D.~M., \& {Guiderdoni}, B. 1993, \mnras, 264, 201, \dodoi{10.1093/mnras/264.1.201}

\bibitem[{{Kemper} {et~al.}(2004){Kemper}, {Vriend}, \& {Tielens}}]{kemper2004}
{Kemper}, F., {Vriend}, W.~J., \& {Tielens}, A.~G.~G.~M. 2004, \apj, 609, 826, \dodoi{10.1086/421339}

\bibitem[{{Kimm} \& {Cen}(2014)}]{kimm2014_SNIa}
{Kimm}, T., \& {Cen}, R. 2014, \apj, 788, 121, \dodoi{10.1088/0004-637X/788/2/121}

\bibitem[{{Kobayashi} {et~al.}(2006){Kobayashi}, {Umeda}, {Nomoto}, {Tominaga}, \& {Ohkubo}}]{kobayashi2006}
{Kobayashi}, C., {Umeda}, H., {Nomoto}, K., {Tominaga}, N., \& {Ohkubo}, T. 2006, \apj, 653, 1145, \dodoi{10.1086/508914}

\bibitem[{{Komatsu} {et~al.}(2011){Komatsu}, {Smith}, {Dunkley}, {Bennett}, {Gold}, {Hinshaw}, {Jarosik}, {Larson}, {Nolta}, {Page}, {Spergel}, {Halpern}, {Hill}, {Kogut}, {Limon}, {Meyer}, {Odegard}, {Tucker}, {Weiland}, {Wollack}, \& {Wright}}]{wmap7}
{Komatsu}, E., {Smith}, K.~M., {Dunkley}, J., {et~al.} 2011, \apjs, 192, 18, \dodoi{10.1088/0067-0049/192/2/18}

\bibitem[{{Kuhn} {et~al.}(2024){Kuhn}, {Guo}, {Martin}, {Bayless}, {Gates}, \& {Puleo}}]{Kuhn_2024}
{Kuhn}, V., {Guo}, Y., {Martin}, A., {et~al.} 2024, \apjl, 968, L15, \dodoi{10.3847/2041-8213/ad43eb}

\bibitem[{{Lacey} {et~al.}(2016){Lacey}, {Baugh}, {Frenk}, {Benson}, {Bower}, {Cole}, {Gonzalez-Perez}, {Helly}, {Lagos}, \& {Mitchell}}]{Lacey_2016}
{Lacey}, C.~G., {Baugh}, C.~M., {Frenk}, C.~S., {et~al.} 2016, \mnras, 462, 3854, \dodoi{10.1093/mnras/stw1888}

\bibitem[{{Lee} {et~al.}(2024){Lee}, {Park}, {Hwang}, \& {Kwon}}]{lee_2024}
{Lee}, J.~H., {Park}, C., {Hwang}, H.~S., \& {Kwon}, M. 2024, \apj, 966, 113, \dodoi{10.3847/1538-4357/ad3448}

\bibitem[{{Leitherer} {et~al.}(1999){Leitherer}, {Schaerer}, {Goldader}, {Delgado}, {Robert}, {Kune}, {de Mello}, {Devost}, \& {Heckman}}]{starburst99}
{Leitherer}, C., {Schaerer}, D., {Goldader}, J.~D., {et~al.} 1999, \apjs, 123, 3, \dodoi{10.1086/313233}

\bibitem[{{Leste} {et~al.}(2024){Leste}, {Willis}, {Canning}, \& {Rennehan}}]{Leste_2024}
{Leste}, O.~K., {Willis}, J.~P., {Canning}, R.~E.~A., \& {Rennehan}, D. 2024, \mnras, 533, 2927, \dodoi{10.1093/mnras/stae1967}

\bibitem[{{Li} {et~al.}(2019){Li}, {Narayanan}, \& {Dav{\'e}}}]{li2019}
{Li}, Q., {Narayanan}, D., \& {Dav{\'e}}, R. 2019, \mnras, 490, 1425, \dodoi{10.1093/mnras/stz2684}

\bibitem[{{Lotz} {et~al.}(2004){Lotz}, {Primack}, \& {Madau}}]{Lotz_2004}
{Lotz}, J.~M., {Primack}, J., \& {Madau}, P. 2004, \aj, 128, 163, \dodoi{10.1086/421849}

\bibitem[{{Lotz} {et~al.}(2008){Lotz}, {Davis}, {Faber}, {Guhathakurta}, {Gwyn}, {Huang}, {Koo}, {Le Floc'h}, {Lin}, {Newman}, {Noeske}, {Papovich}, {Willmer}, {Coil}, {Conselice}, {Cooper}, {Hopkins}, {Metevier}, {Primack}, {Rieke}, \& {Weiner}}]{lotz2008gm20}
{Lotz}, J.~M., {Davis}, M., {Faber}, S.~M., {et~al.} 2008, \apj, 672, 177, \dodoi{10.1086/523659}

\bibitem[{{Maeder} \& {Meynet}(2000)}]{Maeder00}
{Maeder}, A., \& {Meynet}, G. 2000, \aap, 361, 159, \dodoi{10.48550/arXiv.astro-ph/0006405}

\bibitem[{{Martin} {et~al.}(2018){Martin}, {Kaviraj}, {Devriendt}, {Dubois}, \& {Pichon}}]{Martin_2018}
{Martin}, G., {Kaviraj}, S., {Devriendt}, J.~E.~G., {Dubois}, Y., \& {Pichon}, C. 2018, \mnras, 480, 2266, \dodoi{10.1093/mnras/sty1936}

\bibitem[{{Mattsson} \& {Andersen}(2012)}]{Mattsson_2012}
{Mattsson}, L., \& {Andersen}, A.~C. 2012, \mnras, 423, 38, \dodoi{10.1111/j.1365-2966.2012.20574.x}

\bibitem[{{McCluskey} {et~al.}(2024){McCluskey}, {Wetzel}, {Loebman}, {Moreno}, {Faucher-Gigu{\`e}re}, \& {Hopkins}}]{McCluskey_2024}
{McCluskey}, F., {Wetzel}, A., {Loebman}, S.~R., {et~al.} 2024, \mnras, 527, 6926, \dodoi{10.1093/mnras/stad3547}

\bibitem[{{McKinnon} {et~al.}(2017){McKinnon}, {Torrey}, {Vogelsberger}, {Hayward}, \& {Marinacci}}]{McKinnon_2017}
{McKinnon}, R., {Torrey}, P., {Vogelsberger}, M., {Hayward}, C.~C., \& {Marinacci}, F. 2017, \mnras, 468, 1505, \dodoi{10.1093/mnras/stx467}

\bibitem[{{Min} {et~al.}(2007){Min}, {Waters}, {de Koter}, {Hovenier}, {Keller}, \& {Markwick-Kemper}}]{min2007}
{Min}, M., {Waters}, L.~B.~F.~M., {de Koter}, A., {et~al.} 2007, \aap, 462, 667, \dodoi{10.1051/0004-6361:20065436}

\bibitem[{{Mo} {et~al.}(1998){Mo}, {Mao}, \& {White}}]{MoMaoWhite_1998}
{Mo}, H.~J., {Mao}, S., \& {White}, S. D.~M. 1998, \mnras, 295, 319, \dodoi{10.1046/j.1365-8711.1998.01227.x}

\bibitem[{{Moore} {et~al.}(1996){Moore}, {Katz}, {Lake}, {Dressler}, \& {Oemler}}]{Moore_1996}
{Moore}, B., {Katz}, N., {Lake}, G., {Dressler}, A., \& {Oemler}, A. 1996, \nat, 379, 613, \dodoi{10.1038/379613a0}

\bibitem[{{Mu{\~n}oz-Mateos} {et~al.}(2009){Mu{\~n}oz-Mateos}, {Gil de Paz}, {Boissier}, {Zamorano}, {Dale}, {P{\'e}rez-Gonz{\'a}lez}, {Gallego}, {Madore}, {Bendo}, {Thornley}, {Draine}, {Boselli}, {Buat}, {Calzetti}, {Moustakas}, \& {Kennicutt}}]{Mateos_2009}
{Mu{\~n}oz-Mateos}, J.~C., {Gil de Paz}, A., {Boissier}, S., {et~al.} 2009, \apj, 701, 1965, \dodoi{10.1088/0004-637X/701/2/1965}

\bibitem[{{Narayanan} {et~al.}(2010){Narayanan}, {Dey}, {Hayward}, {Cox}, {Bussmann}, {Brodwin}, {Jonsson}, {Hopkins}, {Groves}, {Younger}, \& {Hernquist}}]{Narayanan_2010}
{Narayanan}, D., {Dey}, A., {Hayward}, C.~C., {et~al.} 2010, \mnras, 407, 1701, \dodoi{10.1111/j.1365-2966.2010.16997.x}

\bibitem[{{Negroponte} \& {White}(1983)}]{Negroponte_1983}
{Negroponte}, J., \& {White}, S.~D.~M. 1983, \mnras, 205, 1009, \dodoi{10.1093/mnras/205.4.1009}

\bibitem[{{Popping} \& {P{\'e}roux}(2022)}]{popping_2022}
{Popping}, G., \& {P{\'e}roux}, C. 2022, \mnras, 513, 1531, \dodoi{10.1093/mnras/stac695}

\bibitem[{{Popping} {et~al.}(2017){Popping}, {Somerville}, \& {Galametz}}]{Popping_2017}
{Popping}, G., {Somerville}, R.~S., \& {Galametz}, M. 2017, \mnras, 471, 3152, \dodoi{10.1093/mnras/stx1545}

\bibitem[{{R{\'e}my-Ruyer} {et~al.}(2014){R{\'e}my-Ruyer}, {Madden}, {Galliano}, {Galametz}, {Takeuchi}, {Asano}, {Zhukovska}, {Lebouteiller}, {Cormier}, {Jones}, {Bocchio}, {Baes}, {Bendo}, {Boquien}, {Boselli}, {DeLooze}, {Doublier-Pritchard}, {Hughes}, {Karczewski}, \& {Spinoglio}}]{RemyRuyer_2014}
{R{\'e}my-Ruyer}, A., {Madden}, S.~C., {Galliano}, F., {et~al.} 2014, \aap, 563, A31, \dodoi{10.1051/0004-6361/201322803}

\bibitem[{{Rodriguez-Gomez} {et~al.}(2019){Rodriguez-Gomez}, {Snyder}, {Lotz}, {Nelson}, {Pillepich}, {Springel}, {Genel}, {Weinberger}, {Tacchella}, {Pakmor}, {Torrey}, {Marinacci}, {Vogelsberger}, {Hernquist}, \& {Thilker}}]{Rodriguez_2019}
{Rodriguez-Gomez}, V., {Snyder}, G.~F., {Lotz}, J.~M., {et~al.} 2019, \mnras, 483, 4140, \dodoi{10.1093/mnras/sty3345}

\bibitem[{{Sales} {et~al.}(2012){Sales}, {Navarro}, {Theuns}, {Schaye}, {White}, {Frenk}, {Crain}, \& {Dalla Vecchia}}]{Sales_2012}
{Sales}, L.~V., {Navarro}, J.~F., {Theuns}, T., {et~al.} 2012, \mnras, 423, 1544, \dodoi{10.1111/j.1365-2966.2012.20975.x}

\bibitem[{{Sandage}(1961)}]{Sandage_1961}
{Sandage}, A. 1961, {The Hubble Atlas of Galaxies}

\bibitem[{{Sazonova} {et~al.}(2020){Sazonova}, {Alatalo}, {Lotz}, {Rowlands}, {Snyder}, {Boone}, {Brodwin}, {Hayden}, {Lanz}, {Perlmutter}, \& {Rodriguez-Gomez}}]{Sazonova_2020}
{Sazonova}, E., {Alatalo}, K., {Lotz}, J., {et~al.} 2020, \apj, 899, 85, \dodoi{10.3847/1538-4357/aba42f}

\bibitem[{{Schaller} {et~al.}(1992){Schaller}, {Schaerer}, {Meynet}, \& {Maeder}}]{Schaller92}
{Schaller}, G., {Schaerer}, D., {Meynet}, G., \& {Maeder}, A. 1992, \aaps, 96, 269

\bibitem[{{Schneider} \& {Maiolino}(2024)}]{schneider_2024}
{Schneider}, R., \& {Maiolino}, R. 2024, \aapr, 32, 2, \dodoi{10.1007/s00159-024-00151-2}

\bibitem[{{Scofield} {et~al.}(2025){Scofield}, {Jee}, {Cha}, \& {Park}}]{scofield_2025}
{Scofield}, Z.~P., {Jee}, M.~J., {Cha}, S., \& {Park}, H. 2025, arXiv e-prints, arXiv:2504.08879, \dodoi{10.48550/arXiv.2504.08879}

\bibitem[{{Scoville} {et~al.}(2007){Scoville}, {Aussel}, {Brusa}, {Capak}, {Carollo}, {Elvis}, {Giavalisco}, {Guzzo}, {Hasinger}, {Impey}, {Kneib}, {LeFevre}, {Lilly}, {Mobasher}, {Renzini}, {Rich}, {Sanders}, {Schinnerer}, {Schminovich}, {Shopbell}, {Taniguchi}, \& {Tyson}}]{2007ApJS..172....1S}
{Scoville}, N., {Aussel}, H., {Brusa}, M., {et~al.} 2007, \apjs, 172, 1, \dodoi{10.1086/516585}

\bibitem[{{Somerville} \& {Dav{\'e}}(2015)}]{Somerville_2015}
{Somerville}, R.~S., \& {Dav{\'e}}, R. 2015, \araa, 53, 51, \dodoi{10.1146/annurev-astro-082812-140951}

\bibitem[{{Somerville} {et~al.}(2012){Somerville}, {Gilmore}, {Primack}, \& {Dom{\'\i}nguez}}]{Somerville_2012}
{Somerville}, R.~S., {Gilmore}, R.~C., {Primack}, J.~R., \& {Dom{\'\i}nguez}, A. 2012, \mnras, 423, 1992, \dodoi{10.1111/j.1365-2966.2012.20490.x}

\bibitem[{Springer {et~al.}(2019)Springer, Ofek, Weiss, \& Merten}]{article}
Springer, O., Ofek, E., Weiss, Y., \& Merten, J. 2019, Monthly Notices of the Royal Astronomical Society, 491, \dodoi{10.1093/mnras/stz2991}

\bibitem[{{Sutherland} \& {Dopita}(1993)}]{sutherland_dopita1993}
{Sutherland}, R.~S., \& {Dopita}, M.~A. 1993, \apjs, 88, 253, \dodoi{10.1086/191823}

\bibitem[{{Teyssier}(2002)}]{teyssier_2002}
{Teyssier}, R. 2002, \aap, 385, 337, \dodoi{10.1051/0004-6361:20011817}

\bibitem[{{Thob} {et~al.}(2019){Thob}, {Crain}, {McCarthy}, {Schaller}, {Lagos}, {Schaye}, {Talens}, {James}, {Theuns}, \& {Bower}}]{Thob_2019}
{Thob}, A. C.~R., {Crain}, R.~A., {McCarthy}, I.~G., {et~al.} 2019, \mnras, 485, 972, \dodoi{10.1093/mnras/stz448}

\bibitem[{{Trayford} {et~al.}(2025){Trayford}, {Schaye}, {Correa}, {Ploeckinger}, {Richings}, {Chaikin}, {Schaller}, {Benitez-Llambay}, {Frenk}, \& {Husko}}]{trayford_2025}
{Trayford}, J.~W., {Schaye}, J., {Correa}, C., {et~al.} 2025, arXiv e-prints, arXiv:2505.13056, \dodoi{10.48550/arXiv.2505.13056}

\bibitem[{{Tsai} \& {Mathews}(1995)}]{tsai_1995}
{Tsai}, J.~C., \& {Mathews}, W.~G. 1995, \apj, 448, 84, \dodoi{10.1086/175943}

\bibitem[{{Tweed} {et~al.}(2009){Tweed}, {Devriendt}, {Blaizot}, {Colombi}, \& {Slyz}}]{tweed2009adaptaHOP}
{Tweed}, D., {Devriendt}, J., {Blaizot}, J., {Colombi}, S., \& {Slyz}, A. 2009, \aap, 506, 647, \dodoi{10.1051/0004-6361/200911787}

\bibitem[{{Valentino} {et~al.}(2023){Valentino}, {Brammer}, {Gould}, {Kokorev}, {Fujimoto}, {Jespersen}, {Vijayan}, {Weaver}, {Ito}, {Tanaka}, {Ilbert}, {Magdis}, {Whitaker}, {Faisst}, {Gallazzi}, {Gillman}, {Gim{\'e}nez-Arteaga}, {G{\'o}mez-Guijarro}, {Kubo}, {Heintz}, {Hirschmann}, {Oesch}, {Onodera}, {Rizzo}, {Lee}, {Strait}, \& {Toft}}]{2023ApJ...947...20V}
{Valentino}, F., {Brammer}, G., {Gould}, K. M.~L., {et~al.} 2023, \apj, 947, 20, \dodoi{10.3847/1538-4357/acbefa}

\bibitem[{{van den Bergh}(1960)}]{vandenBergh_1960}
{van den Bergh}, S. 1960, \apj, 131, 558, \dodoi{10.1086/146869}

\bibitem[{{van den Bosch} {et~al.}(2002){van den Bosch}, {Abel}, {Croft}, {Hernquist}, \& {White}}]{Bosch_2002}
{van den Bosch}, F.~C., {Abel}, T., {Croft}, R. A.~C., {Hernquist}, L., \& {White}, S. D.~M. 2002, \apj, 576, 21, \dodoi{10.1086/341619}

\bibitem[{{Ventou} {et~al.}(2017){Ventou}, {Contini}, {Bouch{\'e}}, {Epinat}, {Brinchmann}, {Bacon}, {Inami}, {Lam}, {Drake}, {Garel}, {Michel-Dansac}, {Pello}, {Steinmetz}, {Weilbacher}, {Wisotzki}, \& {Carollo}}]{Ventou_2017}
{Ventou}, E., {Contini}, T., {Bouch{\'e}}, N., {et~al.} 2017, \aap, 608, A9, \dodoi{10.1051/0004-6361/201731586}

\bibitem[{{Wang} {et~al.}(2024){Wang}, {Peng}, {Cappellari}, {Gao}, \& {Mo}}]{Wang_2024}
{Wang}, B., {Peng}, Y., {Cappellari}, M., {Gao}, H., \& {Mo}, H. 2024, \apjl, 973, L29, \dodoi{10.3847/2041-8213/ad772d}

\bibitem[{{Weingartner} \& {Draine}(2001)}]{weingartnerDraine_2001}
{Weingartner}, J.~C., \& {Draine}, B.~T. 2001, \apj, 548, 296, \dodoi{10.1086/318651}

\bibitem[{{White} \& {Rees}(1978)}]{WhiteRees_1978}
{White}, S.~D.~M., \& {Rees}, M.~J. 1978, \mnras, 183, 341, \dodoi{10.1093/mnras/183.3.341}

\bibitem[{{Yan} {et~al.}(2004){Yan}, {Lazarian}, \& {Draine}}]{yan2004}
{Yan}, H., {Lazarian}, A., \& {Draine}, B.~T. 2004, \apj, 616, 895, \dodoi{10.1086/425111}

\bibitem[{{Yao} {et~al.}(2023){Yao}, {Song}, {Kong}, {Fang}, {Zhang}, \& {Chen}}]{yaoyao_2023}
{Yao}, Y., {Song}, J., {Kong}, X., {et~al.} 2023, \apj, 954, 113, \dodoi{10.3847/1538-4357/ace7b5}

\bibitem[{{Zafar} \& {Watson}(2013)}]{Zafar_2013}
{Zafar}, T., \& {Watson}, D. 2013, \aap, 560, A26, \dodoi{10.1051/0004-6361/201321413}

\bibitem[{{Zhukovska}(2014)}]{zhukovska2014}
{Zhukovska}, S. 2014, \aap, 562, A76, \dodoi{10.1051/0004-6361/201322989}

\end{thebibliography}
\bibliographystyle{aasjournal}

\appendix
\twocolumngrid
\counterwithin{figure}{section}

\section{Correlated noise}
\label{sec:correlated_noise}

Following the methodology from \cite{article}, we extracted $m \approx 3\times 10^8$ background-subtracted patches, each of size $7 \times 7$ pixels, with the background being estimated using source-extractor \citep{1996A&AS..117..393B}.
Using these stamps, an empirical covariance matrix can be constructed as follows:
\begin{equation}
    \left(\Sigma_{\mathrm{BG}}\right)_{j,k} = \left(\frac{1}{m-1}\sum_{i=1}^{m}p_i p_i^{\top}\right)_{j,k} \, ,
\end{equation}
where $p_i$ are the vectorized forms of the extracted patches, and each element of $\left(\Sigma_{\rm BG}\right)_{j,k}$ represents the covariance between the $j^{\rm th}$ and $k^{\rm th}$ pixel in the vectorized patches. The matrix square root $L$ of $\Sigma_{\rm BG}$ satisfies $\Sigma_{\rm BG} = L L^{\rm T}$, which allows $L$ to transform uncorrelated i.i.d. standard noise variables into correlated variables that share the covariance structure of $\Sigma_{\rm BG}$:
\begin{align}
    \mathrm{Cov}(L\mathbf{z}) = L\mathrm{Cov}(\mathbf{z})L^{\top} & = L\, I_n\, L^{\top} \nonumber \\ & = L \, L^{\top} = \Sigma_{\rm BG} \, .
\end{align}
Here, $\mathbf{z}$ represents a vector of uncorrelated i.i.d. standard noise variables. The central row of $L$ represents the correlation of the central pixel with itself and all other 48 pixels in a characteristic noise patch. This row can be reshaped into a $7 \times 7$ kernel and convolved with an i.i.d. standard normal noise map, and as a result of this convolution, each pixel in the new noise map becomes the weighted sum of its neighbors in the standard normal noise map. This process effectively introduces the observed correlations and transforms the standard normal noise stamp into a correlated noise stamp that exhibits the covariance properties seen in the observed background noise.

\begin{figure}
\centering
\vspace{2mm}
	\includegraphics[width=\columnwidth]{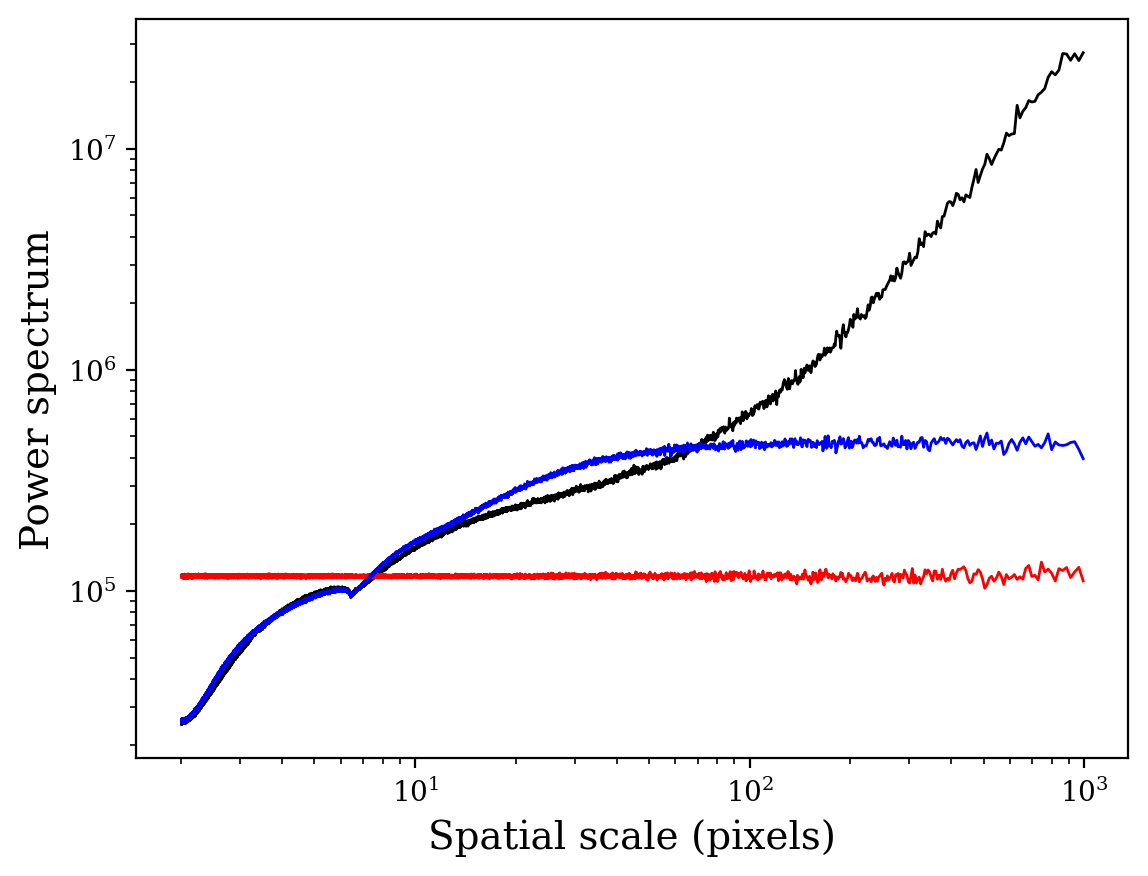}
	\caption{F277W filter power spectral densities for observed noise (black), standard normal noise (red), and correlated noise (blue). At high frequencies, the correlated and observed background noise power spectra are consistent, demonstrating that our method effectively models the small-scale correlations present in the observed background noise.}
	\label{powerspec}
\end{figure}

To evaluate whether the noise characteristics are properly introduced, we investigate the resulting power spectra of the correlated noise maps in a manner similar to \citet{scofield_2025}.
For i.i.d. standard normal noise, a flat power spectrum is expected, showing consistent power across all frequencies. Figure \ref{powerspec} confirms this expectation, displaying a flat profile for standard normal noise. The power spectrum for the correlated background noise aligns well with the observed background noise spectrum at higher frequencies, demonstrating the ability of this noise modeling method to reproduce small-scale correlations introduced by the drizzling process and detector effects. At lower frequencies (corresponding to larger scales), the divergence is anticipated since the method cannot capture or replicate correlations on scales larger than the kernel size. Nevertheless, the method effectively reproduces background noise correlations on scales that are most affected by detector properties and drizzling, and those that could significantly influence galaxy morphology. This ensures the reliability of our morphological analysis in the presence of correlated noise.

\section{Impact of mock observation process on \texorpdfstring{$G\text{\textendash} M_{20}$ distribution}{Importance of the degradation process}}
\label{sec:degrade}

\begin{figure*}
\centering
	\includegraphics[width=\textwidth]{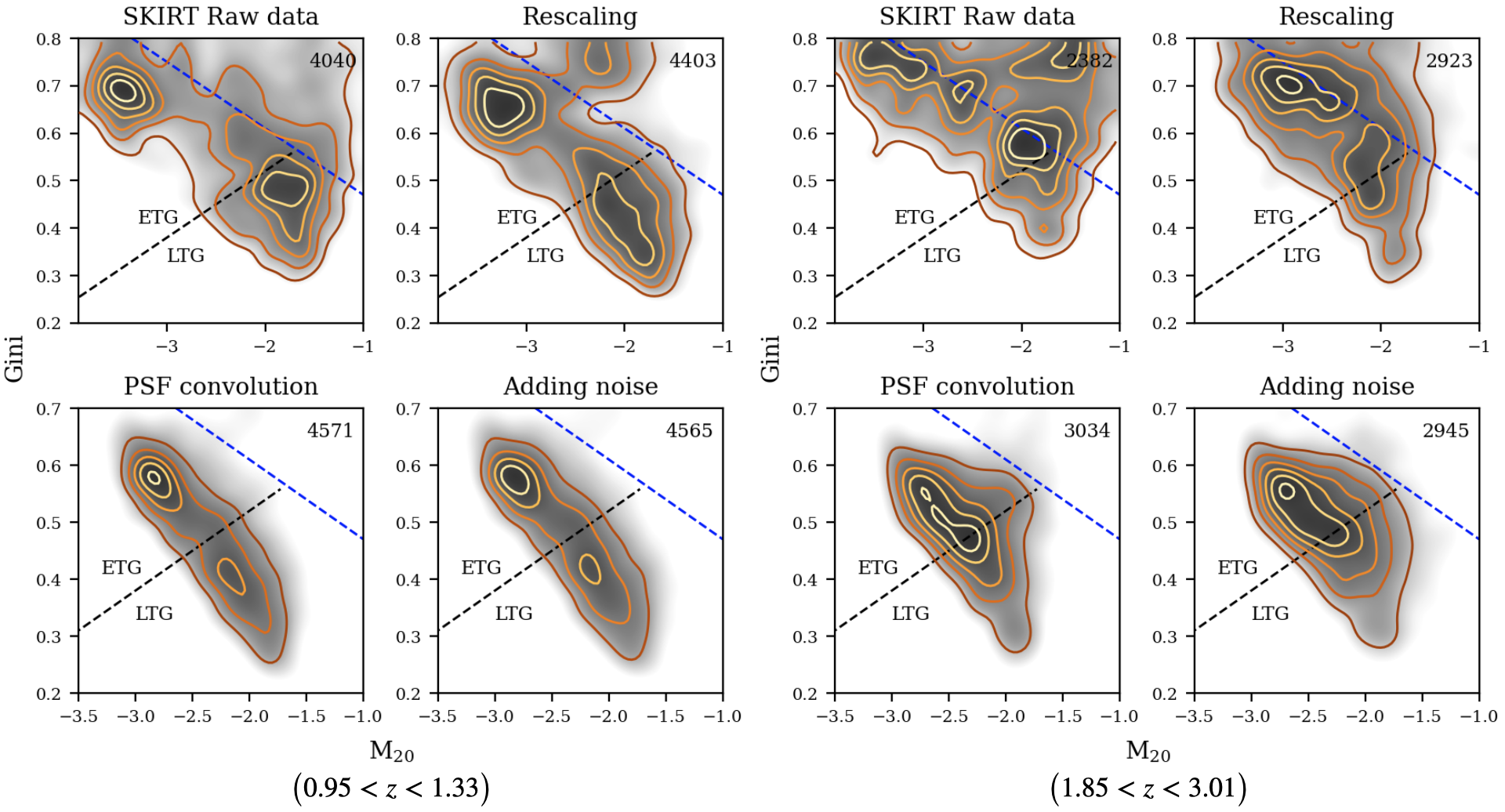}
	\caption{
    The Gini–$M_{20}$ distribution of \NC\ galaxies with the OTF dust model at two different redshift bins: $0.95 < z < 1.33$ (left) and $1.85 < z < 3.01$ (right). Each panel shows the morphological distribution under different mock observation processing steps: raw images (top-left), rescaling (top-right), PSF convolution (bottom-left), and noise addition (bottom-right). The layout and format follow those of Figure~\ref{fig:gmstat}.}
	\label{fig:degrade}
\end{figure*}

We perform a test to evaluate the impact of mock observational processes on the \gm\ distribution of \NC\ galaxies. The results are presented in Figure~\ref{fig:degrade}.

For low-redshift galaxies ($0.95 < z < 1.33$), we find that the PSF convolution step has the most significant impact on the morphological distribution in the \gm\ plane. Before PSF convolution, the distribution is substantially broader, and the separation between the two peaks—corresponding to ETG and LTG—is more pronounced.
The peak position of LTGs remains relatively stable throughout the mock observation steps. In contrast, ETGs exhibit a noticeable shift due to their centrally concentrated light profiles. The PSF convolution redistributes this concentrated flux, leading to a reduction in the Gini coefficient and an increase in $M_{20}$. Consequently, the ETG population moves toward lower Gini and higher $M_{20}$ values after PSF convolution.

A similar trend is observed for high-redshift galaxies ($1.85 < z < 3.01$): PSF convolution again dominates the changes in the \gm\ distribution. However, the effect of noise becomes more prominent compared to the low-redshift case. This is as expected, as high-redshift galaxies typically have lower signal-to-noise ratios, making them more vulnerable to noise-induced uncertainties in morphological measurements.

\end{document}